\documentclass[useAMS,usenatbib,usegraphicx]{mn2e}

\usepackage{graphicx}
\usepackage[latin1]{inputenc}
\usepackage{color}
\usepackage{times}
\usepackage{natbib}
\usepackage{setspace}
\newif\ifAMStwofonts
\AMStwofontstrue
\definecolor{red}{rgb}{1,0.,0.}

\newcommand{\msun}{{\rm M}_\odot}

\newcommand{\gaea}{\sc{gaea}}
\def\lesssim{\lower.5ex\hbox{$\; \buildrel < \over \sim \;$}}
\def\gtrsim{\lower.5ex\hbox{$\; \buildrel > \over \sim \;$}}
\title[CR regulated IMF] {Variations of the stellar initial mass
  function in semi-analytical models II: the impact of Cosmic Ray
  regulation.}  \author[Fontanot et al.]{
  \parbox[t]{\textwidth}{Fabio Fontanot$^1$\thanks{E-mail:
      fontanot@oats.inaf.it}, Gabriella De Lucia$^1$, Lizhi Xie$^1$,
    Michaela Hirschmann$^2$, Gustavo Bruzual$^3$ and St\'ephane
    Charlot$^2$}
    \vspace*{8pt}\\
    $^1$ INAF - Astronomical Observatory of Trieste, via G.B. Tiepolo 11, I-34143 Trieste, Italy \\
    $^2$ Sorbonne Universit\'es, UPMC-CNRS, UMR7095, Institut d'~Astrophysique de Paris, F-75014, Paris, France \\
    $^3$ Instituto de Radioastronom\'\i a y Astrof\'\i sica, UNAM, Campus Morelia, C.P. 58089, Morelia, M\'exico \\
}

\begin{document}
\date{Accepted ... Received ...}

\maketitle

\begin{abstract} 
Recent studies proposed that cosmic rays (CR) are a key ingredient in
setting the conditions for star formation, thanks to their ability to
alter the thermal and chemical state of dense gas in the UV-shielded
cores of molecular clouds. In this paper, we explore their role as
regulators of the stellar initial mass function (IMF) variations,
using the semi-analytic model for GAlaxy Evolution and Assembly
({\gaea}). The new model confirms our previous results obtained using
the integrated galaxy-wide IMF (IGIMF) theory: both variable IMF
models reproduce the observed increase of $\alpha$-enhancement as a
function of stellar mass and the measured $z=0$ excess of dynamical
mass-to-light ratios with respect to photometric estimates assuming a
universal IMF. We focus here on the mismatch between the
photometrically-derived ($M^{\rm app}_{\star}$) and intrinsic
($M_{\star}$) stellar masses, by analysing in detail the evolution of
model galaxies with different values of $M_{\star}/M^{\rm
  app}_{\star}$. We find that galaxies with small deviations
(i.e. formally consistent with a universal IMF hypothesis) are
characterized by more extended star formation histories and live in
less massive haloes with respect to the bulk of the galaxy
population. While the IGIMF theory does not change significantly the
mean evolution of model galaxies with respect to the reference model,
a CR-regulated IMF implies shorter star formation histories and higher
peaks of star formation for objects more massive than $10^{10.5}
\msun$. However, we also show that it is difficult to unveil this
behaviour from observations, as the key physical quantities are
typically derived assuming a universal IMF.
\end{abstract}

\begin{keywords}
  galaxies: formation - galaxies: evolution - galaxies: abundances -
  galaxies: fundamental parameters - galaxies: stellar content
\end{keywords}

\section{Introduction}\label{sec:intro}                                       
Star formation represents a key, yet not fully understood, physical
mechanism governing galaxy evolution. In a star formation episode, one
of the fundamental parameters of the associated stellar population is
the so-called Initial Mass Function (IMF), i.e. the number of stars
formed per stellar mass bin. The IMF has long been seen as a nearly
invariant property of star forming regions, mainly due to the
remarkable consistency of its shape measured in the local
neighbourhood (but apparently not in the densest regions of the
Galactic Centre, see e.g. \citealt{Klessen07}). Different functional
forms have been proposed to describe the IMF \citep[among the most
  popular]{Salpeter55, Kroupa01, Chabrier03}. These differ mainly for
the abundance of stars at the low-mass end up to the brown dwarf
regime. Unfortunately, direct measurements of the IMF via star counts
is accessible only for local star forming regions (typically within
our own Milky Way or its closest satellites), and we are thus forced
to resort to the analysis of the integrated light of stellar
populations in external galaxies. Indirect evidence against a
universal\footnote{Throughout this paper, we will neglect the (small)
  differences between the \citet{Chabrier03} IMF and a
  broken-power-law formulation (see \citealt{Kroupa13} for a more
  detailed comparison between these different functional forms).}  IMF
has been found for different galaxy populations, such as late-type
\citep{Gunawardhana11}, early-type \citep{Ferreras13} and dwarf
galaxies \citep{McWilliam13}. These claims have been recently
confirmed by dynamical and spectroscopic studies suggesting relevant
deviations from a universal IMF as a function of galaxy stellar mass
or velocity dispersion. For a sample of early-type galaxies in the
ATLAS$^{3D}$ survey, \citet{Cappellari12} contrasted integral-field
maps of stellar kinematics and optical imaging against dynamical
models, including the contribution of both stellar and Dark Matter
(DM) components. Their results show a systematic excess of the
dynamical mass-to-light ratios with respect to photometric estimates
based on a universal IMF. This excess increases with galaxy velocity
dispersion ($\sigma$). The interpretation of this result in terms of
IMF variations is not straightforward. A possible solution is given by
the assumption that massive galaxies are characterized by a
``top-heavy'' IMF, which implies a larger fraction of massive
(short-lived) stars and a decrease in the total light expected at late
times (at fixed stellar mass), with respect to a universal IMF
assumption. Conversely, a ``bottom-heavy'' solution is also plausible:
it corresponds to a larger fraction of low-mass stars and an increase
of the stellar mass at fixed light, with respect to a universal
IMF. \citet[see also \citealt{Tortora16}]{Cappellari12} were unable to
distinguish between the two scenarios. Compatible results were
obtained by \citet[see also \citealt{Ferreras13} and
  \citealt{Spiniello12}]{ConroyvanDokkum12} using a spectrophotometric
approach. Their analysis uses spectral features in galaxy spectra
sensitive to the stellar effective temperature and surface gravity,
and thus able to constrain the ratio between low-mass and giant stars,
which gives information on the IMF shape. In particular, they compare
high-resolution spectra of a sample of compact early type galaxies
against libraries of synthetic models obtained by varying the IMF
slopes at the high-mass and low-mass ends. The pre-selection of
compact elliptical galaxies allows them to assume that the observed
$\sigma$ within the effective radius is dominated by stellar kinematic
(neglecting the DM contribution). They interpret their best fit
solutions in favour of an increasingly bottom-heavy IMF with
increasing $\sigma$ and/or stellar mass.

From a theoretical point of view, it has long been argued that the IMF
could and should vary as function of local properties of the
interstellar medium (ISM) in star forming regions. A number of
theoretical models (see e.g. \citealt{Klessen05, WeidnerKroupa05,
  HennebelleChabrier08, Papadopoulos10, Hopkins12}) have analyzed the
role of small-scale physical properties of the ISM in star forming
regions in setting the IMF shape, and predict a wide range of possible
IMF shapes. They differ mainly in the modelling of the physical
processes that govern the fragmentation and formation of stars of
different mass from the parent molecular cloud (MC). As we already
mentioned, it is currently impossible to study the IMF shape directly
(i.e. via star counts), but we typically constrain it using the
integrated light of composite stellar populations in external
galaxies. This complicates the picture, as our analysis of a composite
stellar population depends on our understanding of the simple stellar
populations which form at different times and locations \citep[see
  e.g.][]{KroupaWeidner03, Kroupa13}. As many physical properties of
galaxies change both spatially and temporally, we expect the redshift
evolution of the IMF shape to play a fundamental role in setting the
physical properties of galaxies. These considerations suggests that
available data can provide us only instantaneous (often galaxy-wide)
information on a ``mean'' or ``effective'' IMF. Dedicated theoretical
modelling is crucial to test the proposed models, following the build
up of the composite stellar populations of different galaxy classes
and comparing their synthetic properties with observational
constraints. Using this approach, it will be possible to simulate
observational samples, prepare mock catalogues and characterize
selection effects. This is particularly relevant, given the current
claims of possible inconsistencies between the dynamical and
spectroscopic approach \citep[see e.g.][]{Smith14}.
 
In a previous paper, \citet[F17a hereafter, see also
  \citealt{Fontanot14}]{Fontanot17a} we tested the impact of the
so-called Integrated Galaxy-wide stellar Initial Mass Function (IGIMF)
theory, first proposed by \citet[][see also
  \citealt{WeidnerKroupa05}]{KroupaWeidner03}, on predictions from our
GAlaxy Evolution and Assembly ({\gaea}) model. This model is
particularly well designed for this kind of studies as it features (i)
a detailed chemical enrichment model \citep{DeLucia14}, following the
evolution of several species and abundance ratios and (ii) an improved
treatment for stellar feedback, gas ejection and re-accretion
\citep{Hirschmann16} that reproduces the redshift evolution of the
galaxy stellar mass function over a wide range of stellar masses and
redshifts \citep{Fontanot17b}. The IGIMF theory directly relates the
mean shape of the IMF to the instantaneous star formation rate (SFR)
of each galaxy, so that it is well suited for theoretical models not
resolving the details of star forming regions in galaxies. The IGIMF
variation as a function of SFR corresponds to a change of the
high-stellar mass slope, with stronger SFR episodes corresponding to
top-heavier IMFs.

In this work, we will study the implications of an independent
approach, first proposed by \citet{Papadopoulos10} and later developed
in \citet[P11 hereafter]{Papadopoulos11}. The role of Cosmic Rays
(CR), associated with supernovae (SNe) explosions and stellar winds
around young stars, in the process of star formation has been long
debated in the literature \citet[see e.g.][]{Krumholz14}, given their
higher efficiency in penetrating the inner dense, UV-shielded regions
of MCs, with respect to UV-photons. As an intriguing possibility, PP11
postulate that the local CRs energy density could indeed regulate the
minimum temperature, ionization state and chemical composition of
these dense MC cores and, ultimately, their characteristic Jeans
mass. This is a quite different assumption with respect to the IGIMF
theory, that postulates that individual molecular clouds are
characterized by a canonical (Kroupa-like) IMF. Nonetheless, it is
worth stressing that the effect of CRs should be implicitly accounted
for in the IGIMF theory, as it is based on an empirical calibration of
the stellar IMF and its dependence on local properties \citep{Marks12,
  Kroupa13}. The predicted evolution of the IMF shape is however
sensibly different from the IGIMF theory: in the PP11 formalism, the
main evolution is seen in the characteristic mass (i.e. the position
of the knee of the IMF). These two models represent two independent
ways of varying the IMF shape: a comparison of their predictions will
thus provide insight on their different implications in terms of star
formation history and mass assembly.

This paper is organized as follows. In Section~\ref{sec:igimf} we will
outline the basics of the CR theory as presented in PP11
and~\citet{Papadopoulos13}. We will then describe its semi-analytic
implementation in Section~\ref{sec:models}. We will present our
results and compare them with the IGIMF expectations in
Section~\ref{sec:results}. Finally, we will summarise and discuss our
results in Section~\ref{sec:final}.
                                                                              
\section{Cosmic rays as regulators of star formation}\label{sec:igimf}
In order to estimate the IMF variations as a function of the CR
density, we follow the approach described in PP11 (see also
\citealt{Papadopoulos13}). PP1 assume that CRs associated with SNe and
stellar winds provide an extra heating source for star-forming
molecular clouds (MCs), affecting their ionization and chemical
state. It is possible to estimate the minimum temperature
($T_{\rm k}$) of the ISM \citep[see e.g.][and references
  herein]{Elmegreen08} and the thermal state of the gas
\citep[e.g.][]{Jasche07} including the CR heating term ($\Gamma_{\rm
  CR}$) in a thermal balance equation \citep{Goldsmith01}:

\begin{equation}
\Gamma_{\rm CR} = \Lambda_{\rm line}(T_{\rm k}) + \Lambda_{\rm gd}(T_{\rm k})
\label{eq:Gfirst}
\end{equation}

\noindent
where $\Lambda_{\rm line}$ and $\Lambda_{\rm gd}$ represent the
contribution to gas cooling from molecular lines and gas-dust
interactions\footnote{Explicit expressions for each cooling term can
  be found in Appendix A of PP11}, and $\Gamma_{\rm CR}$
depends on the MC gas core density ($n_{\rm H2}$) and on the CR
ionization rate ($\gamma_{\rm CR}$).

\begin{equation}
\Gamma_{\rm CR} \propto n_{\rm H2} \gamma_{\rm CR}
\label{eq:Gsecond}
\end{equation}

\noindent
Following PP11, we assume that $\gamma_{\rm CR}$ is proportional to
the CR energy density $U_{\rm CR}$. Using a simple analytic model
based on a fixed chemical composition, PP11 showed that the minimum
temperature in dense MC gas cores is much higher in galaxies with SFR
densities larger than our own Milky Way. A more general solution for
the chemical and thermal equations regulating the CR-dominated ISM
requires the modelling of the evolution of ISM chemistry, which is
also strongly dependent on $\gamma_{\rm CR}$; this solution can be
determined numerically \citep[see e.g.][]{Thi09}. PP11 discussed the
evolution of the gas temperature as a function of CR energy densities
$T_{\rm k}(\Gamma_{\rm CR})$ and MC gas core densities. The
characteristic Jeans mass of young stars ($M^{\star}_{\rm J}$) can be
written as:

\begin{displaymath}
  M^{\star}_{\rm J} = \left[ k_{\rm B} \frac{T_{\rm k}(\Gamma_{\rm CR})}{G \mu m_{\rm H2}} \right]^{3/2} \rho_{\rm c}^{-1/2} 
\end{displaymath}
\begin{equation}
  \, \, = 0.9 \left( \frac{T_{\rm k}(\Gamma_{\rm CR})}{10 K} \right)^{3/2} \left( \frac{n_{\rm H2}}{10^4 {\rm cm}^{-3}} \right)^{-1/2} \msun.
  \label{eq:mjc}
\end{equation}

\noindent
In PP11 there is no attempt to estimate a global IMF shape (i.e. to be
associated with the whole sta forming galaxy): their results hold for
individual star forming MCs, given their gas and CR densities. In
order to translate these results into a prescriptions to be
implemented in {\gaea}, we thus need some additional assumptions.

\begin{figure*}
  \centerline{ \includegraphics[width=9cm]{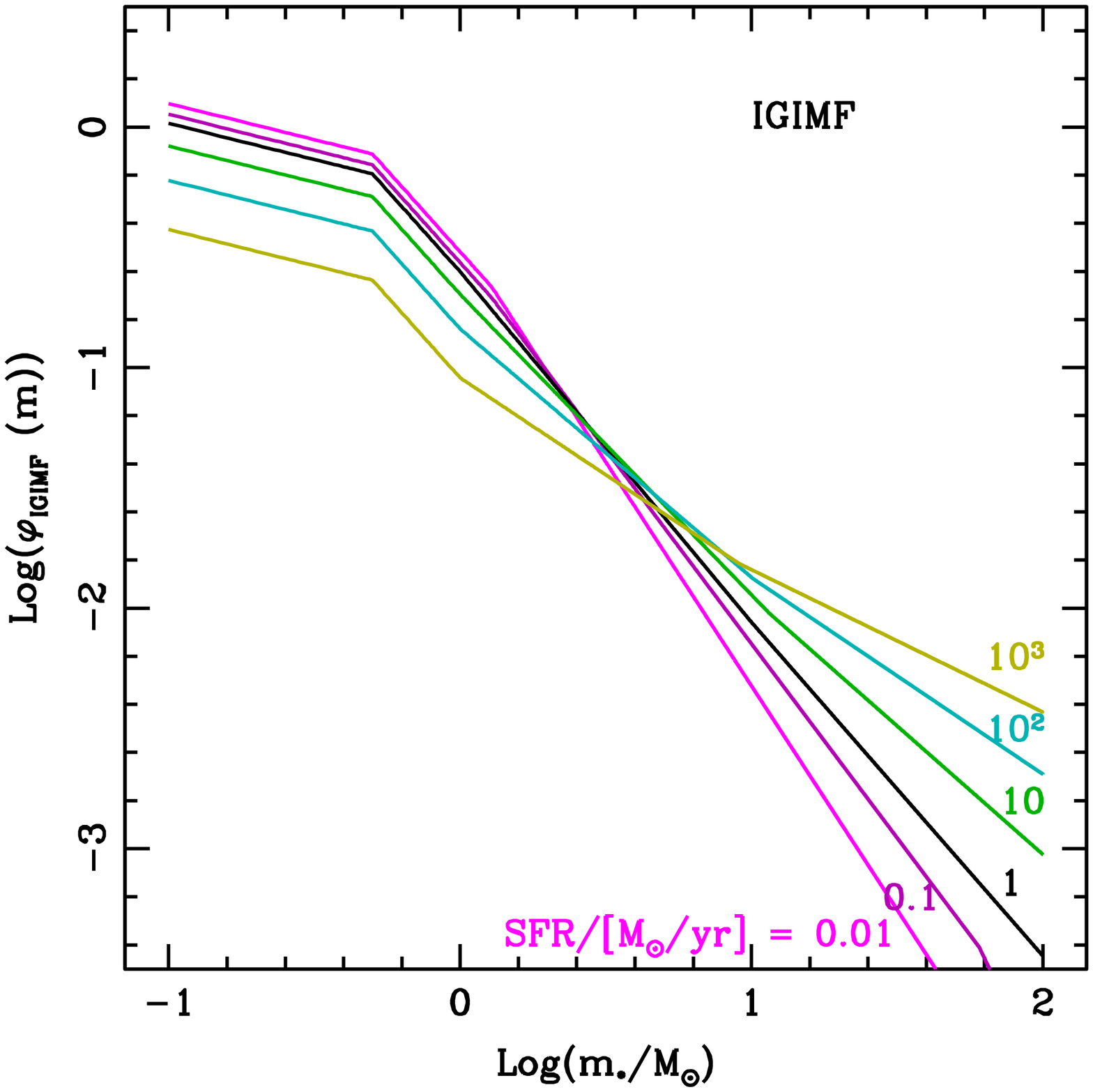}
    \includegraphics[width=9cm]{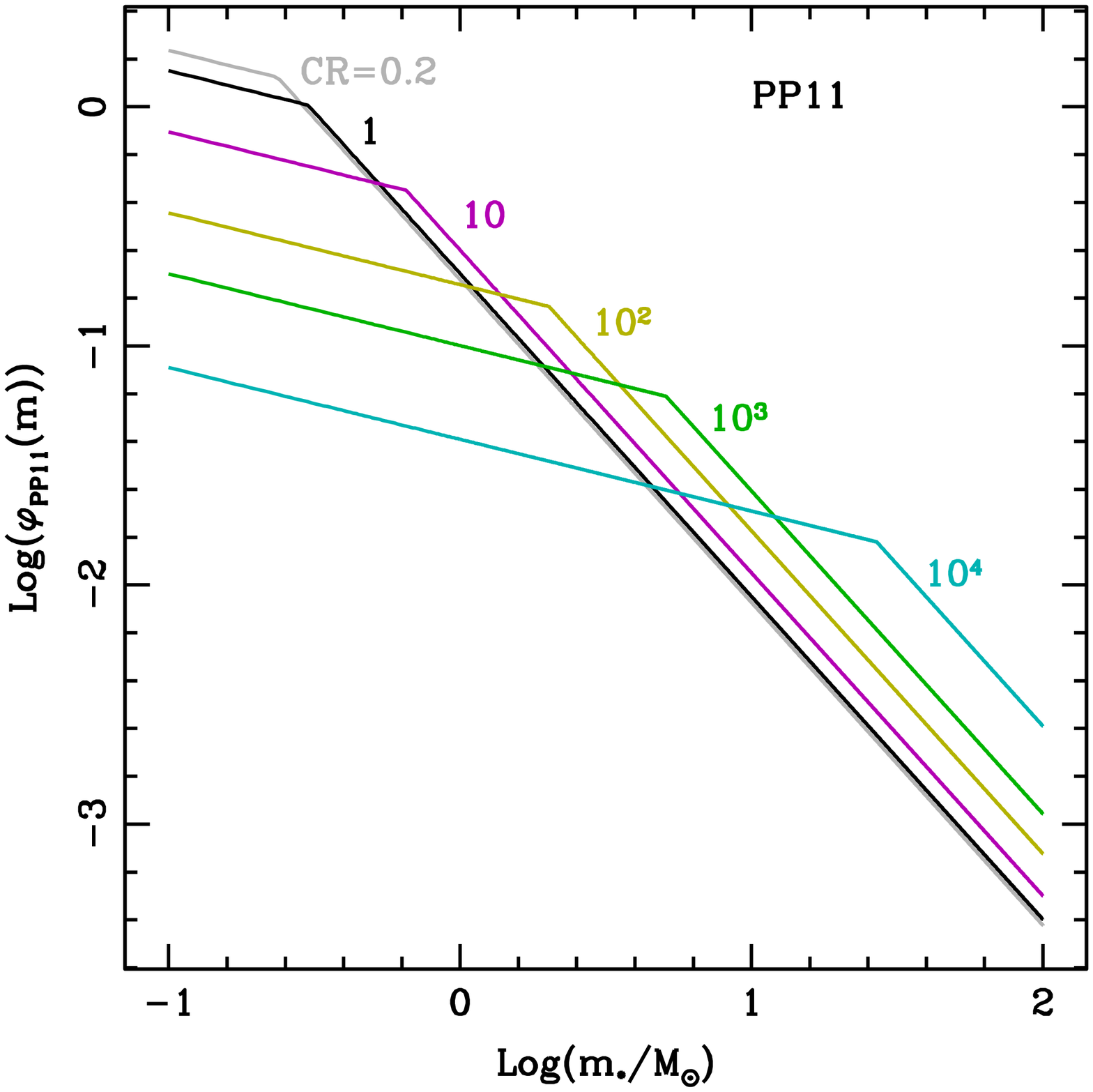} }
  \caption{Variable IMF scenarios. {\it Left panel:} evolution of the
    IMF shape according to PP11. Labels close to lines report the
    corresponding cosmic ray densities (normalized to the Milky Way
    value, $CR = U_{\rm CR}/U_{\rm MW}$). {\it Right panel:} IMF
    variation as predicted by the IGIMF theory. Labels close to the
    lines report the reference SFRs (see also Fig.~1 in F17a). In both
    panels, each IMF is normalized to 1 $\msun$ in the stellar mass
    interval 0.1-100 $\msun$.}\label{fig:pp_imf}
\end{figure*}

First of all we assume that the overall IMF shape is well described by
a broken power law (see e.g. \citealt{Kroupa01}), characterized by two
slopes:

\begin{equation}
\varphi_\star(m) =  \left\{
\begin{array}{ll}
(\frac{m}{m_{\rm low}})^{-\alpha_1} & m_{\rm low} \le m < m_{\rm break} \\
(\frac{m_0}{m_{\rm low}})^{-\alpha_1} (\frac{m}{m_0})^{-\alpha_2} & m_{\rm break} \le m < m_{\rm max} \\
\end{array}
\right.
\label{eq:kroimf}
\end{equation}

\noindent
where $m_{\rm low}=0.1 \msun$, $m_{\rm max}=100 \msun$,
$\alpha_1=1.3$, $\alpha_2=2.35$. Such a general shape is motivated by
local observations of individual clouds, and it agrees well with
theoretical calculations based on fragmentation of giant molecular
clouds \citep[see e.g.][and reference
  herein]{HennebelleChabrier08}. In our models, we will assume that
the high- and low-mass ends of the IMF have fixed slopes, and that
only the characteristic mass of young stars, i.e. the knee $m_{\rm
  break}$ of the IMF is affected by the CR density. At a given CR
density (i.e. $T_{\rm k}(\Gamma_{\rm CR})$), we assume that $m_{\rm
  break}$ corresponds to $M^{\star}_{\rm J}$ at $n(H2)=10^5$
cm$^{-3}$, a representative density for typical molecular clouds
(Papadopoulos, private communication). In detail, we adopt the
numerical solutions by PP11 (their Fig.~4, see also
Tab.~\ref{tab:mbval}), where $M^{\star}_{\rm J}$ is computed as a
function of $n(H2)$ for 6 different values of CR energy density
(normalized to the MW value, $U_{\rm CR}/U_{\rm MW}$). We approximate
the CR energy density in our models galaxies by the SFR surface
density ($\Sigma_{\rm SFR}$). Following PP11, we define 6 different IMF
shapes (Fig.~\ref{fig:pp_imf}) assuming:

\begin{equation}
\frac{U_{\rm CR}}{U_{\rm MW}} = \frac{\Sigma_{\rm SFR}}{\Sigma_{\rm MW}}
\end{equation}

\noindent
where $\Sigma_{\rm MW} = 10^{-3} \msun$ yr$^{-1}$ kpc$^{-2}$ is the
estimated SFR surface density for the MW disc. It is important to keep
in mind that this description of the IMF variation has not been tested
yet against resolved stellar populations, such as globular clusters
and/or young open clusters. In fact, these observations suggest an
invariant break mass. The main line of evidence in favor of cosmic ray
regulated IMF variations come from the properties of starburst
galaxies \citep{Papadopoulos13, Romano17}. CRs may also shift the
average stellar mass in a population rather than only the break mass
and may thus partially or wholly be responsible for the changes in the
stellar IMF deduced by \citet{Marks12}.
\begin{table}
  \caption{Analytic description for the IMF variations}
  \label{tab:mbval}
  \renewcommand{\footnoterule}{}
  \centering
  \begin{tabular}{ccccccccc}
    \hline
    $U_{\rm CR}/U_{\rm MW}$ & $m_{\rm low}$ & $\alpha_1$ & $m_{\rm break}$ & $\alpha_2$ & $m_{\rm max}$ \\
     & [M$_\odot$] & & [M$_\odot$] & & [M$_\odot$] \\
    \hline
     0.2    &   0.1  & 1.3 &  0.25 &  2.35 &   100 \\
     1.0    &   0.1  & 1.3 &  0.30 &  2.35 &   100 \\
     10     &   0.1  & 1.3 &  0.65 &  2.35 &   100 \\
     10$^2$ &   0.1  & 1.3 &  2.   &  2.35 &   100 \\
     10$^3$ &   0.1  & 1.3 &  5.   &  2.35 &   100 \\
     10$^4$ &   0.1  & 1.3 &  27.  &  2.35 &   100 \\
    \hline
  \end{tabular}
\end{table}
%
\section{Semi-analytic Model}\label{sec:models}
In order to assess the influence of the proposed variable IMF model on
the chemical and physical properties of galaxy populations, we
implement it in the GAlaxy Evolution and Assembly ({\gaea})
semi-analytic model (SAM). This method follows the evolution of galaxy
properties along cosmic epochs by modelling the network of relevant
physical mechanisms acting on the baryonic component of dark
matter haloes, whose hierarchical evolution (traced using N-body
cosmological simulations) represents the gravitational backbone of the
model. These baryonic processes (which include the cooling and heating
of baryonic gas, star formation, accretion of gas onto Super-Massive
Black Holes and the related feedback processes) are described using
approximated analytic and/or numerical prescriptions that are
observationally and/or theoretically motivated: this results in a very
flexible tool to predict galaxy properties for galaxy samples in
cosmological volumes.

Our {\gaea} model provides significant improvements with respect to
previous versions of the code \citep{DeLuciaBlaizot07}. In this
section we will quickly sketch the relevant changes implemented for
the present analysis, and we refer the interested reader to the
original papers for more details. {\gaea} implements a detailed
treatment of chemical enrichment, described in \citet{DeLucia14},
which accounts explicitly for the different lifetimes of stars of
different mass and their peculiar enrichment patterns. The
differential lifetimes and yields for individual chemical species from
Asymptotic Giant Branch (AGB) stars, Type II SNe and Type Ia SNe are
explicitly tracked by the code.

A second relevant change in the code corresponds to the updated
modelling of stellar feedback discussed in \citet{Hirschmann16}. In
that paper, we show that some form of preventive or ejective feedback
is needed to reproduce the observed space density evolution of
galaxies below the knee of the stellar mass function. Our reference
feedback scheme (H16F in the following) corresponds to a prescription
combining (i) gas reheating and energy injection schemes mutuated from
high-resolution hydrodynamical simulations \citep[i.e.]{Muratov15},
with (ii) a gas ejection scheme based on energy conservation arguments
\citet{Guo11} and (iii) a timescale of gas re-incorporation that
depends on halo mass \citep{Henriques13}. In \citet{Hirschmann16}, we
show that this model is also able to reproduce the gas fractions and
mass metallicity relations at $z<3$, while \citet{Fontanot17b}
demonstrated that the agreement of the H16F run with the evolution of
the stellar mass function and cosmic star formation rate extends to
$z\sim7-10$. It is worth stressing that the H16F feedback
prescriptions do not include any explicit dependence on the number of
SNe. However, in a variable IMF scenario, the fraction of SNe per unit
stellar mass formed ($f_{\rm SN}$) is not constant. In order to
account for this effect we compute $f_{\rm SN}$ for the 6 discrete IMF
shapes considered in this paper (see below). For each SF event, we
then rescale the feedback efficiencies with the ratio between the
$f_{\rm SN}$ of the chosen IMF and the MW-like IMF (i.e. $U_{\rm
  CR}/U_{\rm MW}=1$). In the following, we will refer to predictions
of this model as the PP11 run\footnote{We explicitly tested that
  results from our reference PP11 run are qualitatively similar to
  those of a run implementing the same variable IMF model, but without
  any $f_{\rm SN}$ scaling.}.

We compare these predictions with already published results from F17a
(namely the ``High-$\alpha_{\rm SF}$ model in F17a - IGIMF run in this
paper). In F17a we introduced the code modifications needed to deal
with a variable IMF. Those include the generalisation of look-up
tables used by the code to track the enrichment of individual chemical
elements and the energy given by each Single Stellar Population at
different cosmic epochs. As in F17a, we construct photometric tables
corresponding to the binned IMFs in Tab.~\ref{tab:mbval}, using an
updated version of the \citet{Bruzual03} model (Bruzual \& Charlot, in
preparation). This includes a prescription for the evolution of
thermally pulsing AGB stars \citep{Marigo08}. The treatment of dust
extinction is the same as in \citet{DeLuciaBlaizot07}.

In addition to these changes, the variable IMF runs we develop in this
paper require further improvements in the code. Star formation in
{\gaea} takes place in galaxy discs, with only a minor contribution
from collisional starbursts in mergers \citep{Fontanot13}. In the PP11
run, we take advantage of the updated modelling of disc sizes for
{\gaea} described in \citet[see also \citealt{Guo11}]{Xie17}. While
gas and stellar discs grow in mass, the code tracks their angular
momentum evolution following the mass and energy exchanges between the
different galaxy components (halo, disc and bulge). \citet{Xie17}
showed that the updated model predicts gas and stellar disks larger
than the original H16F run, especially for more massive galaxies, in
better agreement with observational measurements. Overall, the updated
disc size model does not affect dramatically the predictions of the
H16F run, and in particular statistical properties like the stellar
mass functions and the mass-metallicity relations.

In this paper, we will use {\gaea} predictions based on the merger
trees extracted from the Millennium Simulation \citep{Springel05},
i.e. a $\Lambda$CDM concordance cosmology, with parameters\footnote{As
  shown in previous papers \citep[e.g.][]{Wang08, Guo13}, we do not
  expect the mismatch of these cosmological parameters with respect to
  the most recent measurements \citep{Planck_cosmpar} to change our
  main conclusions.} derived from WMAP1 (i.e. $\Omega_\Lambda=0.75$,
$\Omega_m=0.25$, $\Omega_b=0.045$, $n=1$, $\sigma_8=0.9$, $H_0=73 \,
{\rm km/s/Mpc}$). We will consider three different models, i.e. the
original H16F model (as in \citealt{Hirschmann16}), a model
implementing the IGIMF theory, and the new model based on the PP11
approach. All these runs implement the same feedback scheme (as in
H16F). However, the changes in galaxy evolution induced by a variable
IMF approach are such that a recalibration of the key {\gaea}
parameters is needed. As for the IGIMF runs, we choose to recalibrate
our model by requiring it to reproduce the evolution of
multiwavelength luminosity functions. In fact, we cannot use the
evolution of the galaxy stellar mass function (GSMF) and/or cosmic SFR
as calibration set, since the observational estimates for these
physical quantities are derived under the assumption of a universal
IMF. We give more details on the recalibration process in
Appendix~\ref{app:recal}. The values of the relevant parameters are
compared to the H16F and IGIMF parameters in Tab.~\ref{tab:runs}.

In next sections, we will compare {\gaea} predictions with
observational constraints involving physical properties of galaxies
(like $M_{\star}$ or SFRs). These are usually derived from broadband
photometry (spectral energy distribution fitting or colour scalings)
or spectroscopy under the assumption of a universal/invariant IMF. In
order to perform a proper comparison with our model predictions, we
follow the same approach as in F17a and define an {\it apparent} -
Chabrier (2003) IMF equivalent - stellar mass ($M_\star^{\rm app}$)
using a mass-to-light vs colour relation calibrated on synthetic
magnitudes (see F17a and Appendix~\ref{app:phys} for more details). In
brief, $M_\star^{\rm app}$ represents the stellar mass that an
observer would derive from the synthetic photometry under the
assumption of a universal IMF, while $M_{\star}$ is the intrinsic
stellar mass predicted by the model. Both $M_{\star}$ and
$M_\star^{\rm app}$ account for the mass of stellar remnants.

\section{Results}\label{sec:results}
\begin{table}
  \caption{Parameter values adopted for the runs considered in this
    study (parameters are defined in Appendix~\ref{app:phys})}.
  \label{tab:runs}
  \renewcommand{\footnoterule}{}
  \centering
  \begin{tabular}{cccc}
    \hline
    Parameter & H16F & IGIMF & PP11 \\
    \hline
    $\alpha_{\rm SF}$            & 0.03 & 0.19  & 0.08 \\
    $\epsilon_{\rm reheat}$       & 0.3  & 0.575 & 0.30 \\
    $\epsilon_{\rm eject}$        & 0.1  & 0.12  & 0.04 \\
    $\gamma_{\rm reinc}$          & 1.0  & 1.0   & 0.50 \\
    $\kappa_{\rm radio} / 10^{-5}$ & 1.0  & 1.78  & 1.18 \\
    \hline
  \end{tabular}
\end{table}

In this section we will compare predictions based on the PP11
approach, with our previous runs implementing either the IGIMF
framework or the canonical IMF (H16F). We aim to explore common trends
between the variable IMF scenarios and highlight discrepant
predictions that can be used to disentangle between the different
scenarios. The predicted evolution of key physical and chemical
properties for model galaxies in the PP11 run confirm the main
conclusions by F17a in the framework of the IGIMF theory. Namely, the
inclusion of a variable IMF provides a viable explanation for some
recent and puzzling observational results and long standing
problems. These models are able to recover, in particular, the
observed relation between the [$\alpha$/Fe] enrichment and the stellar
mass of elliptical galaxies, and the correct trend for the $z=0$
mass-metallicity relation. \citet{DeLucia17} discuss the difficulties
in reproducing both relations {\it at the same time} in theoretical
models of galaxy formation (both SAMs and hydrodynamical simulations)
using a universal IMF. Moreover, both IMF variation models are able to
qualitatively reproduce the observational evidence in favour of a
non-universal IMF, coming from dynamical analysis (see
e.g. \citealt{Cappellari12}) or spectral synthesis models (see
e.g. \citealt{ConroyvanDokkum12} or \citealt{LaBarbera17}). The
prediction of the IGIMF and PP11 models are quite similar for these
observables, and hardly distinguishable within the errorbars. The
interested reader will find the plots corresponding to these
predictions in Appendix~\ref{app:phys}, together with a more detailed
discussion. In the following we will focus on the main differences we
find between the two models and the reference H16F run.

\begin{figure}
  \centerline{ \includegraphics[width=9cm]{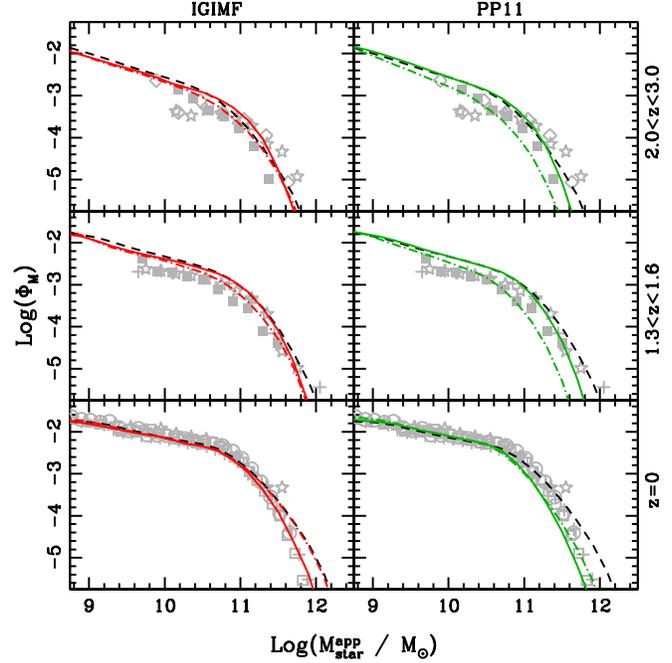} }
  \caption{Redshift evolution of the galaxy stellar mass function. In
    each panel, the black long-dashed line refers to the predictions
    of the H16F model; the solid lines to the GSMF as a function of
    $M^{\rm app}_{\star}$ and the dot-dashed lines to the GSMF as a
    function of the intrinsic galaxy stellar mass $M_{\star}$. Grey
    points show a collection of observational measurements as in
    \citet[see detailed references
      herein]{Fontanot09b}. }\label{fig:mfs}
\end{figure}

In Fig.~\ref{fig:mfs}, we compare the GSMFs at different redshifts in
the H16F, IGIMF and PP11 runs. In each panel, coloured solid lines
refer to the GSMFs as a function of $M^{\rm app}_{\star}$. Using
photometrically-equivalent stellar masses, the GSMF evolution is quite
similar between H16F, IGIMF and PP11 runs (black dashed lines show the
evolution in the H16F run). This is a very interesting conclusion,
that highlights the intrinsic difficulty of using the stellar mass
function as a discriminant of galaxy evolution, in a varying IMF
context. However, the situation is dramatically different when
considering the predicted GSMFs as a function of the intrinsic stellar
mass $M_{\star}$ (dot-dashed coloured lines). As shown in F17a, in the
IGIMF framework the differences are small for most redshifts, with the
main deviations seen at low redshifts and high masses. In particular
the GSMFs based on $M^{\rm app}_{\star}$ are in good agreement with
the $z=0$ GSMF, thus easing the discrepancy between the intrinsic GSMF
and observational data. The picture is completely different for the
PP11 approach: the intrinsic GSMFs correspond to space densities
systematically smaller than those obtained with $M^{\rm app}_{\rm
  star}$. The systematic difference holds for the highest redshift
probed (i.e. $z\sim7$), and {\it reduces} at $z<1$. This mismatch is a
clear indication that the mass assembly is radically different in the
PP11 run with respect to H16F (and IGIMF) runs. In the following
sections we will focus on this aspect.

\subsection{Comparison between PP11 and IGIMF theory.}\label{sec:comp}
\begin{figure}
  \centerline{ \includegraphics[width=9cm]{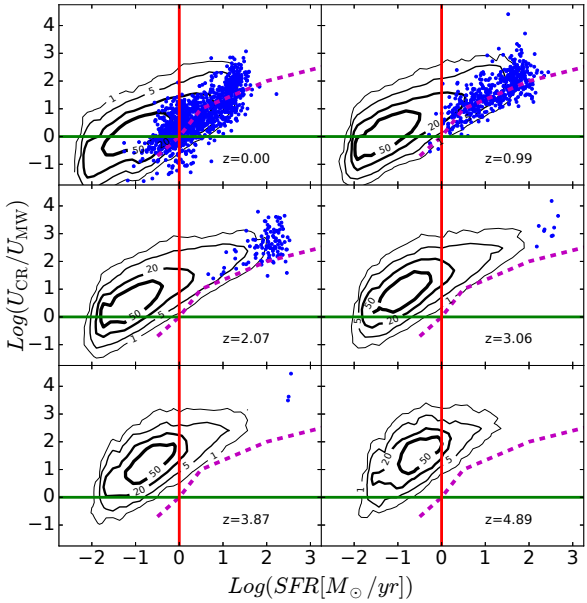} }
  \caption{Distribution of model galaxies in the SFR vs $U_{\rm
      CR}/U_{\rm MW}$ space in the PP11 run, at different
    redshift. Contours correspond to number densities for the global
    population, while blue dots mark the position of $M_{\rm
      star}>10^{11} \msun$ galaxies. Red vertical and green horizontal
    lines mark the values corresponding to a MW-like IMF in IGIMF and
    PP11 formalism, respectively. Magenta dashed line connects points
    corresponding to IMFs with similar $f_{\rm SN}$ in the two
    formalisms (see text for more details).}\label{fig:cuts}
\end{figure}
\begin{figure*}
  \centerline{ \includegraphics[width=18cm]{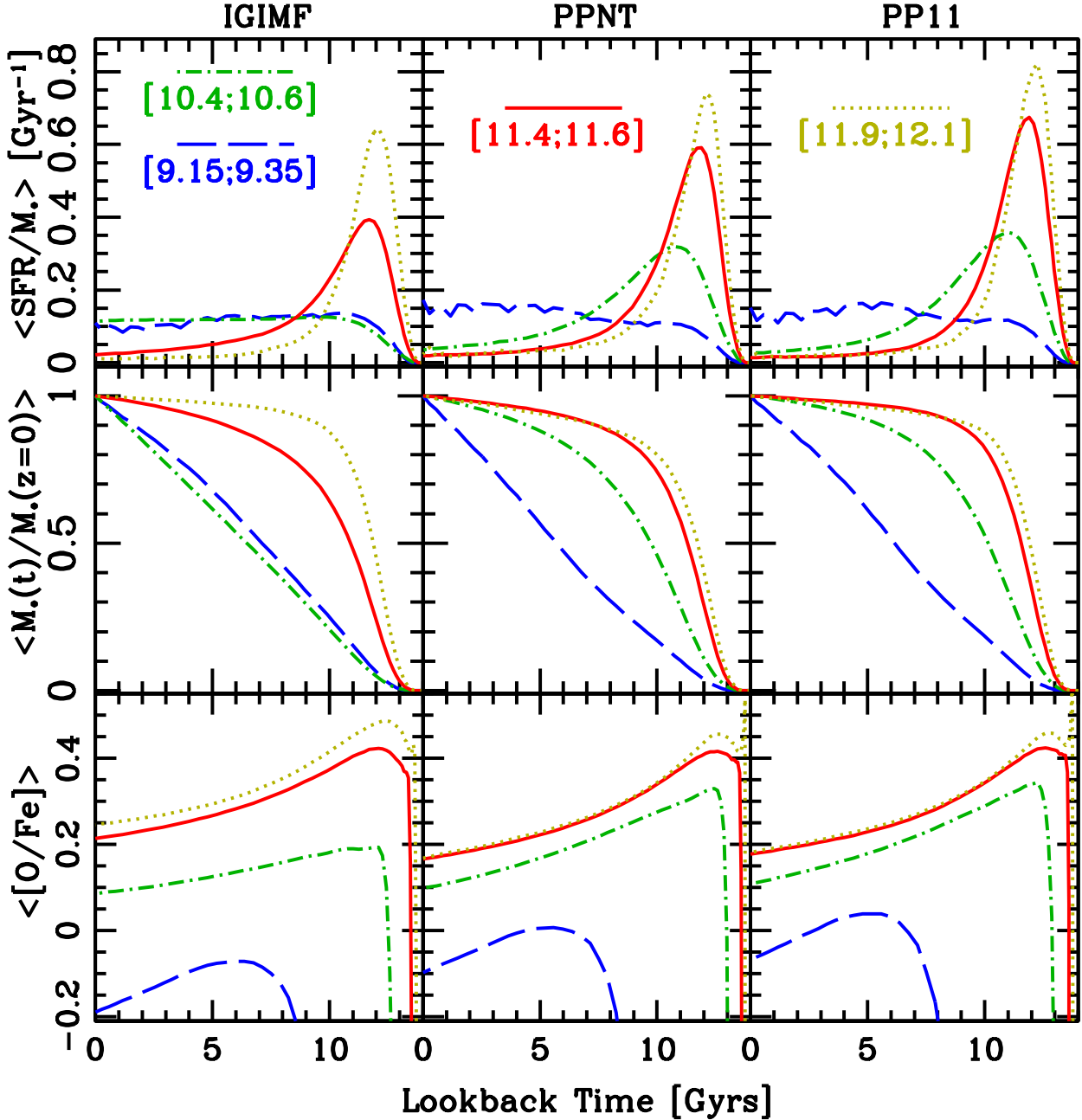} }
  \caption{Mean normalized star formation histories (upper panels),
    cumulative mass assembly (middle panels) and evolution of the
    [O/Fe] ratio (lower panels) for galaxies in the different
    logarithmic stellar mass intervals (see caption). The predictions
    relative to the IGIMF and PP11 models are shown in the right-hand
    and left-hand columns, respectively. The middle column shows the
    prediction for a model (PPNT) implementing the PP11 model, but
    using the same parameters as in H16F.}\label{fig:sfh}
\end{figure*}
The differences in the GSMF evolution can be better understood if we
consider the distribution of {\gaea} model galaxies in the SFR -
$U_{\rm CR}/U_{\rm mw}$ space. Fig.~\ref{fig:cuts} shows this
distribution for galaxies from the PP11 run and highlights the
differences between the two IMF scenarios. Contour plots mark the
number density levels corresponding to 1, 5, 20 and 50 percent. Blue
dots in Fig.~\ref{fig:cuts} show the position of massive galaxies
(i.e. $M_{\star}>10^{11} \msun$). In each panel, the red vertical and
green horizontal lines correspond to the SFR and $U_{\rm CR}/U_{\rm
  mw}$ values typical of a MW-like galaxy, i.e. to a canonical IMF in
the PP11 and in the IGIMF theory, respectively. Above the green lines
and to the right of red lines model galaxies are characterized by
``Top-Heavy'' IMFs in the corresponding variable IMF scenario, and
viceversa. The figure show that most galaxies in the IGIMF run form
stars with an IMF that is comparable and/or bottom-heavier than the MW
one, while massive galaxies tend to have a top-heavy IMF, especially
at high-redshifts. Deviations from a MW-like IMF are more marked in
the PP11 model (i.e. there is a stronger evolution along the
y-axis). In this case, the fraction of model galaxies forming stars
with a MW-like (or bottom-heavy IMF) is decreasing with increasing
redshift. A more quantitative comparison between the two models is
complicated by the different evolution predicted for the IMF
shape. Indeed, a ``Top-Heavy'' IMF in the PP11 runs is not the same as
in the IGIMF framework. To better compare the two scenarios, we use
the fraction of SN per unit of stellar mass formed ( $f_{\rm SN}$). It
should be noticed, however, that given the different shape of the IMF,
similar values of $f_{\rm SN}$ correspond to different ratios between
SNIa and SNII. We compute $f_{\rm SN}$ for the 6 IMF shapes considered
in this work and for the 21 ones in F17a. The dashed magenta line in
Fig.~\ref{fig:cuts} connects the points corresponding to similar
values of $f_{\rm SN}$ in the two runs. If a model galaxy lies above
this line, it is forming stars with an IMF that is top-heavier in the
PP11 run than in IGIMF theory: this is the case for most sources at
high-redshift. At $z \lesssim 1$ massive model galaxies are more
evenly distributed around the magenta line, i.e. the two approaches
predict comparable IMF shapes. Fig.~\ref{fig:cuts} clearly shows that
the PP11 implementation corresponds to larger deviations from the
universal IMF hypothesis than in the IGIMF theory.

\subsection{Mass assembly history.}

Given the previous discussion, it is not surprising that the largest
discrepancies between the two variable IMF models are seen in the mass
assembly history of model galaxies. In order to highlight this effect,
in Fig.~\ref{fig:sfh} we show the mean star formation histories,
cumulative mass assembly and evolution of the [O/Fe] ratio for four
different $z=0$ mass bins (corresponding to $M_{\star} \sim 10^{12}$,
$10^{11.5}$, $10^{10.5}$ and $10^{9.25} \msun$). Fig.~\ref{fig:sfh}
can be compared with a similar figure in F17a, where we showed that
the star formation and mass assembly histories in the H16F and IGIMF
runs are quite similar. The situation is completely different for the
PP11 model (right column). The evolution of galaxies more massive than
$M_{\star} \sim 10^{10.5} \msun$ is somehow accelerated, with a higher
peak of star formation at high redshift and a shorter star formation
time-scales with respect to their analogous in the IGIMF model (left
column), thus enhancing the more rapid formation of more massive
galaxies in hierarchical models of galaxy evolution
\citep{DeLucia06}. Model galaxies in the smallest mass bin show a
reversal of this trend. While in the IGIMF (and H16F) runs they are
characterized by flat star formation histories, in the PP11 run they
clearly show a rising SFR at late times. This behaviour is driven by
the implementation of the PP11 formalism, and only mildly affected by
the recalibration of the model. In order to test this, we run a model
realization (PPNT) using the PP11 scenario with the same {\gaea}
parameters as in H16F: the corresponding results are shown in
Fig.~\ref{fig:sfh} (middle column) and clearly demonstrate that the
change in mass assembly history for massive galaxies is driven by the
IMF variation.

\subsection{Mass excess distributions.}
\begin{figure}
  \centerline{ \includegraphics[width=9cm]{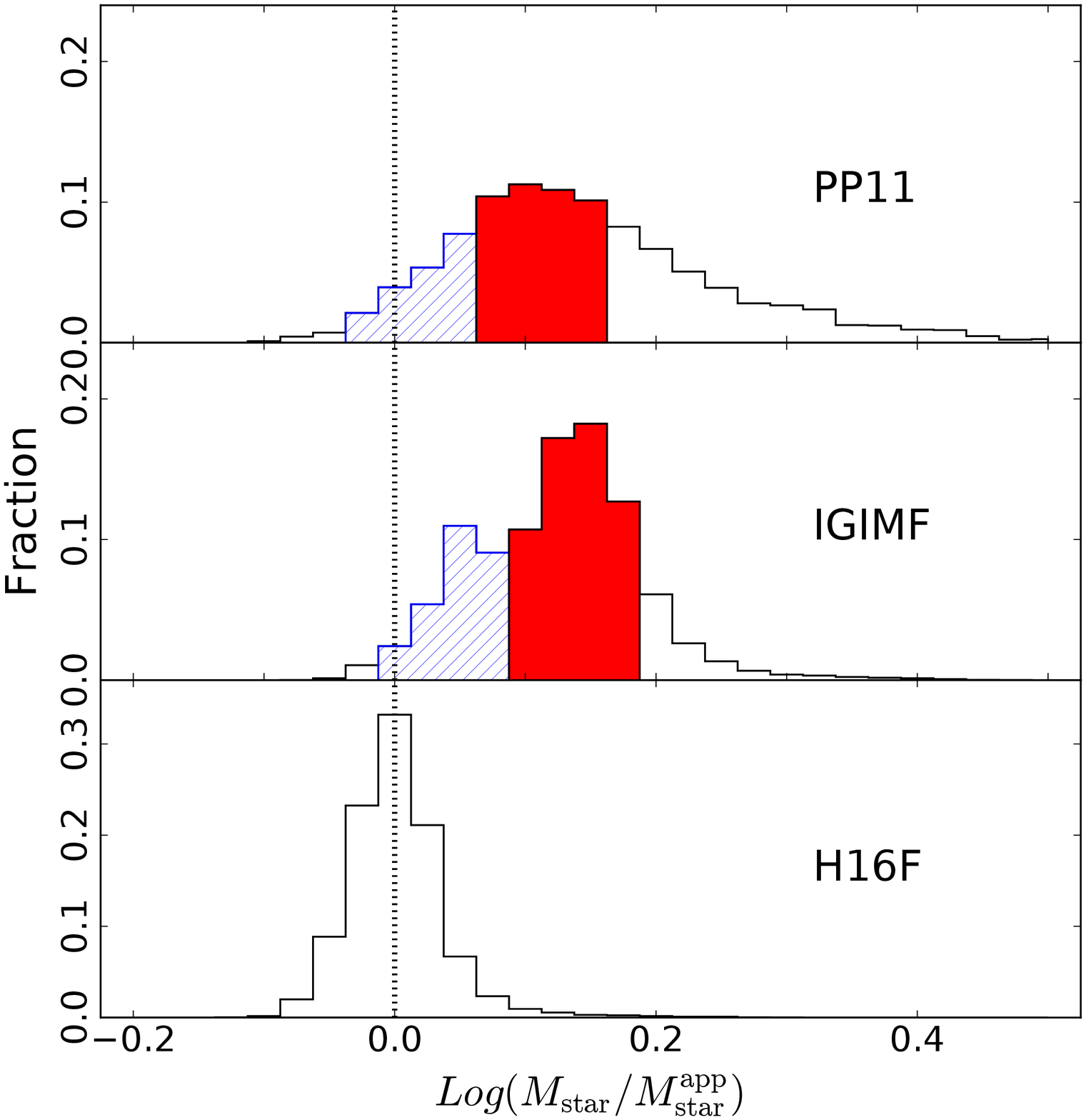} }
  \caption{Mass excess distributions for $M_{\star} \sim 10^{11.5}
    \msun$ galaxies in the PP11, IGIMF and H16F runs. The red shaded
    area (subsample H) includes galaxies around the peak of the
    distribution, while the blue hatched region (subsample N) selects
    galaxies with small mass excesses (i.e. compatible with the
    hypothesis of a universal IMF).}\label{fig:slicehisto}
\end{figure}
\begin{figure}
  \centerline{ \includegraphics[width=9cm]{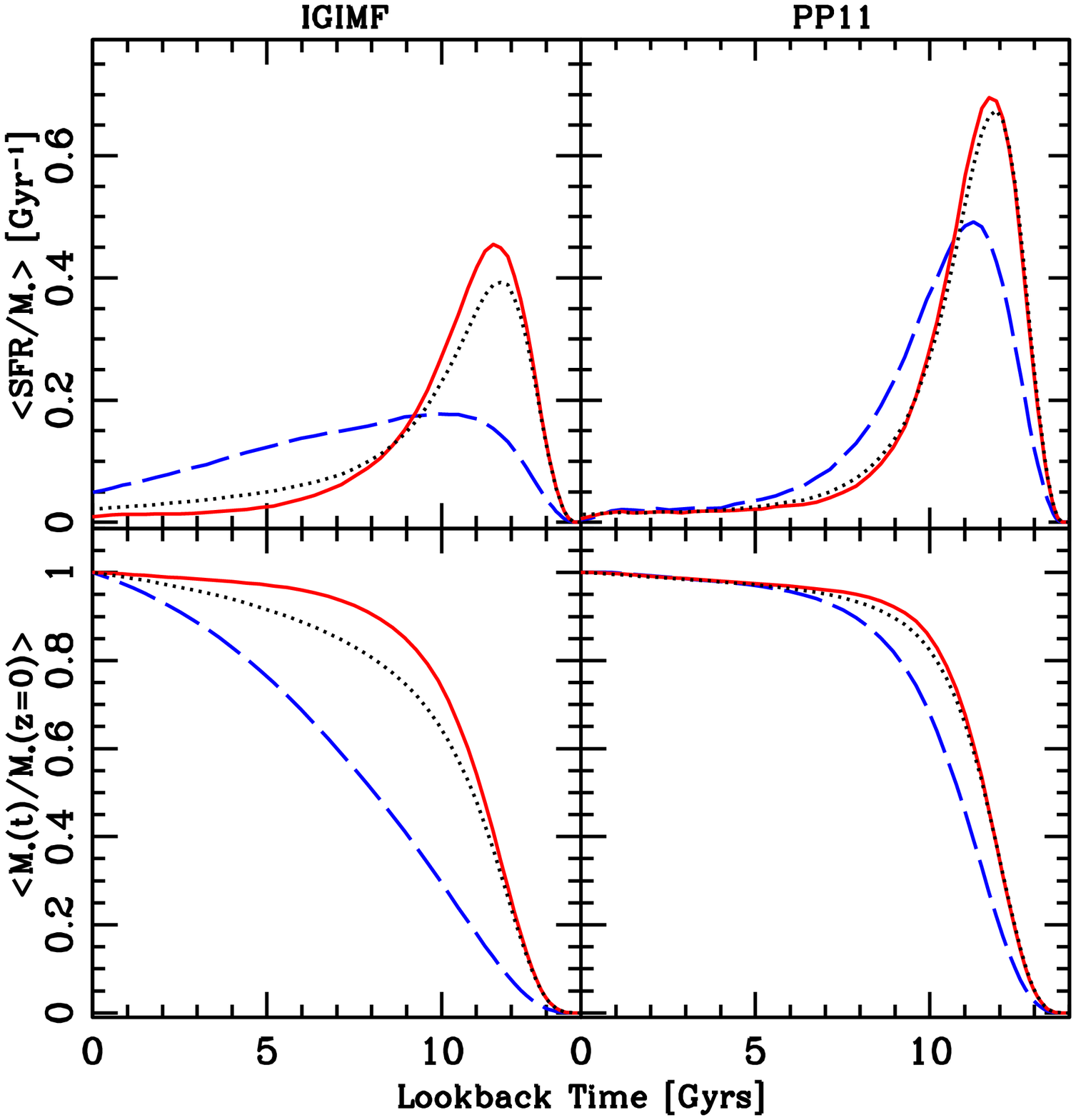} }
  \caption{Mean normalized star formation histories (top panels) and
    cumulative mass assembly (bottom panels) for $M_{\star} \sim
    10^{11.5} \msun$ galaxies. Black dotted, blue dashed and red solid
    lines refer to the total sample, to subsample N and subsample H,
    respectively.}\label{fig:slice}
\end{figure}
As we mentioned earlier, one of the key successes of runs implementing
IMF variations (either assuming the IGIMF or the PP11 theory) is the
prediction of a discrepancy between the intrinsic $M_{\star}$ and the
apparent $M^{\rm app}_{\star}$, growing at increasing $M_{\star}$. If
we interpret the former as a dynamical mass estimate and the latter as
a photometric mass estimate, we can associate the $M_{\star}/M^{\rm
  app}_{\star}$ ratio to the so-called ``mass excess'' found in
dynamical and spectroscopic studies. At high-$M_{\star}$, the
dispersion of the predicted relation is wide enough that a
non-negligible fraction of model galaxies has $M_{\star}/M^{\rm
  app}_{\star} \sim 1$, i.e. they do not show a mass excess. This
result is interesting by itself, as it potentially explains
contrasting results obtained for massive lensed galaxies
\citep{Smith15, Leier16}, that are found to have mass-to-light ratios
consistent with a universal IMF. In order to understand the origin of
this galaxy population in our models, we consider the mass excess
distribution for model galaxies in a slice corresponding to the mass
range $11.4<\log(M_{\star}/\msun)<11.6$
(Fig.~\ref{fig:slicehisto}). The position of the peak of the
distribution is clearly different with respect to the H16F run
(featuring a universal IMF) in both variable IMF scenarios. There are
also differences between the PP11 and IGIMF runs, with the latter
showing a broader distribution (and more model galaxies compatible
with the universal IMF scenario). We use the information in
Fig.~\ref{fig:slicehisto} to define two subsamples of model galaxies
in the PP11 and IGIMF runs. The first one (subsample H - red shaded
area) includes galaxies around the peak of the distribution, while the
second one (subsample N - blue hatched region) corresponds to the tail
of the distribution, i.e. galaxies with small mass excesses and whose
global properties are compatible with the hypothesis of a universal
IMF. These two subsamples are defined using different
$M_{\star}/M^{\rm app}_{\star}$ ratios, due to the different shape of
the parent distribution in the two runs.

\begin{figure}
  \centerline{ \includegraphics[width=9cm]{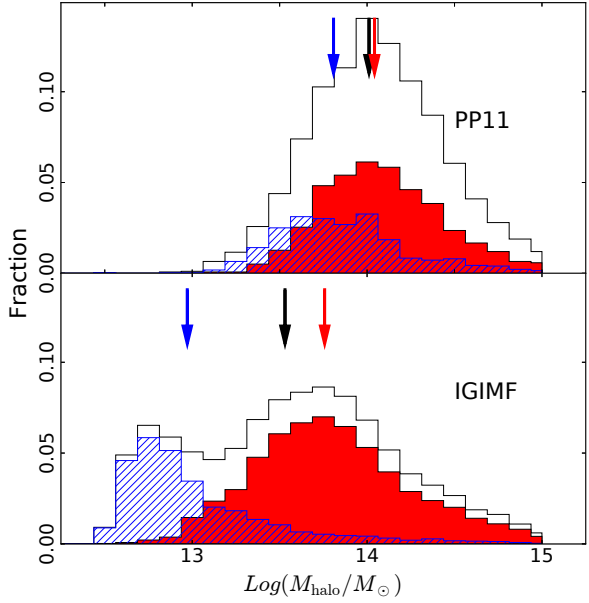} }
  \caption{Parent halo mass distributions for $M_{\star} \sim
    10^{11.5} \msun$ galaxies in the PP11, IGIMF and H16F runs. Red
    shaded and blue hatched regions refer to subsample H and N, as in
    Fig.~\ref{fig:slicehisto}. Black, red and blue arrows mark the
    mean halo mass for the whole sample and subsample H and N,
    respectively}\label{fig:mhalo}
\end{figure}

We then repeat the same analysis as in Fig.~\ref{fig:sfh} for the
galaxies in the two subsamples. In Fig.~\ref{fig:slice} we compare the
resulting mean normalized star formation histories and cumulative mass
assembly with those of the parent $M_{\star} \sim 10^{11.5} \msun$
population (black dotted lines). Subsample N galaxies (blue dashed
lines) are characterized by a redshift evolution which is quite
different with respect to both the total sample and subsample H (red
solid line). In particular they tend to form the bulk of their
population at later times, and over a longer star formation timescale
with respect to the total population. This different evolutionary
history relates to the different environment these galaxies live
in. If we consider the distribution of their parent DM halo masses
(Fig.~\ref{fig:mhalo}), we clearly find a systematic difference
between subsample N and H: galaxies with small (or no) mass excess
tend to live in less massive haloes with respect to those close to the
peak of the mass excess distribution. The black, blue and red arrows
in Fig.~\ref{fig:mhalo} mark the mean halo mass associated with the
global population, subsample H and subsample N, respectively. This
environmental signature is particularly evident in the IGIMF run,
where the two distributions are well separated, while in the PP11 run
there is a relevant overlap between the two populations.

\section{Discussion \& Conclusions}\label{sec:final}
In this paper, we present a version of the {\gaea} semi-analytic
model, including a model for a variable IMF based on PP11, who
characterized the impact of the CR energy density on the properties of
star forming regions. We assume that the main effect on the IMF shape
is an evolution of the characteristic mass (i.e. the knee of the
function), that increases with increasing SFR surface density (we use
this quantity as a proxy for CR energy density). We contrast
predictions from this model with those from our previous work, based
on the IGIMF framework \citep{WeidnerKroupa05}. This alternative
scenario predicts a top-heavier IMF with increasing SFR. This
represents a further development for {\gaea}, which is now able to
handle two different prescriptions for IMF variations, driven by
different physical properties of model galaxies (although SFRs and SFR
surface densities are correlated quantities, as we show and discuss in
Sec.~\ref{sec:comp}). In this paper we discuss similarities and
differences between the two IMF variation scenarios, and we discuss
the implications in terms of galaxy evolution and assembly.

The PP11 and IGIMF formalisms predict different shapes for the
IMF. The latter model is based on a deterministic derivation of the
IMF, based on the assumption of a universal (Kroupa-like) IMF in
individual clouds. The instantaneous galaxy-wide SFR is the main
driver of the variation, and mainly affects the slope(s) of the IMF at
the high-mass end. In the PP11 approach, the star formation conditions
in dense UV-shielded MC cores are determined by the CR energy
density. This sets the minimum temperature of the gaseous phase and
modifies the Jeans mass estimate. The CR density can be connected to
the star formation surface density, which we use as the physical
parameter governing the IMF variation in this scenario. It is
important to realise that in the PP11 model the IMF variation is a
local process, taking place inside MCs, and possibly varying across
different locations within galaxies. This is a clear difference with
respect to the IGIMF theory, that offers an empirical derivation of
the galaxy integrated IMF, based on and compatible with the observed
properties of individual star clusters \citep{Yan17}.

In this work, we model the IMF variation driven by CR regulation as a
change in the knee of the IMF, keeping the high-mass and low-mass
slopes fixed. Moreover, we compute the IMF variation in the entire
star forming disc (i.e. we are using the total SFR surface density as
the leading parameter). Further studies, with hydrodinamical
simulations able to resolve the internal structure of galaxies, will
allow us to understand the effect of these variable IMF scenarios on
the radial properties of galactic discs.

The variable IMF models considered here provide consistent predictions
for the evolution of key physical quantities, especially when these
quantities are computed using an {\it apparent} - Chabrier IMF
equivalent - stellar mass from synthetic photometry. In particular,
the PP11 run is able to reproduce the [$\alpha$/Fe]-stellar mass
relation \citep[see e.g.][]{Johansson12} and the mass-to-light
(stellar mass) excess inferred for local elliptical galaxy
samples. These results are in agreement with those found in our
previous work, in the framework of the IGIMF model. Moreover, we show
that these observational constraints are not able to disentangle
between the two scenarios. Nonetheless, our results show the
robustness of the predictions of galaxy formation models against
different variable IMF frameworks.

In particular, in F17a, we showed that in the GAEA run implementing
the IGIMF theory the mass assembly and star formation histories of
model galaxies are hardly changed with respect to the H16F run
adopting a universal IMF. The situation is different for the PP11
model. In this run, the intrinsic mass assembly of model galaxies is
dramatically altered. Above $10^{10.5} \msun$, mean star formation
histories become more peaked, and the peak moves at higher redshift
with respect to the H16F predictions.  Albeit these differences are
obvious and significant when considering model results, they are
difficult to unveil from photometric data. Indeed, the stellar mass
functions derived from photometrically derived $M_\star^{\rm app}$ are
indistinguishable between H16F, IGIMF and PP11 runs.

As in the case of the IGIMF model, PP11 predictions are in qualitative
agreement with the results presented in \citet{Conroy13},
i.e. with an excess of the intrinsic stellar mass with respect to the
photometric expectation. This mass excess is increasingly larger at
increasing stellar mass. Our runs contain a significant number of
model galaxies characterized by synthetic photometry and mass-to-light
ratios compatible with a universal IMF scenario. This finding can
explain results based on the lensing analysis of a limited samples of
early-type galaxies \citep{Smith15, Leier16}. We study in detail the
properties of this subsample, and we show that these model galaxies
assemble on a longer timescale with respect to the bulk of the
population, and preferentially live in less massive DM halos.

It is worth stressing, that our analysis is based on the mismatch
between synthetic photometry and intrinsic physical properties of
model galaxies. In particular, the mass excess in our runs is not due
to an intrinsic ``bottom-heavier'' IMF in massive galaxies. This
contrasts with the interpretation of \citet{ConroyvanDokkum12} and
\citet{LaBarbera13}, based on the analysis of spectral features
sensitive to the giants to low-mass stars ratio. In future work, we
plan to deepen this aspect by generating synthetic spectra based on
the star formation histories extracted from our variable IMF runs. We
will then analyze these synthetic spectra using the same approach used
in the observed samples, in order to better characterize the level of
disagreement between theoretical predictions and reconstructed IMF
shapes.

Another interesting approach is based on the study of the relative
abundances of CNO isotopes in the ISM. \citet{Romano17} showed that
the cosmic evolution of isotopes of these chemical species is
extremely sensitive to the fraction of massive stars (AGB stars and
novae) and thus to the IMF shape. They argued that in order to explain
the isotope ratios (mainly $^{12}C/^{13}C$ and $^{16}O/^{18}O$)
observed in starbursts, an IMF skewed towards high-stellar masses is
required. Line ratios of isotopologues (i.e. molecules that differ in
their isotopic composition) have practical advantages to study IMF
variations: most of them are accessible in the submmillimetre regime
and are mostly insensitive to dust obscuration. Current facilities
like ALMA already extend the accessible redshift range up to
$z\sim3$. All these attempts represent the necessary next steps in
theoretical studies of IMF variations, and will be at the centre of
our future work.

\section*{Acknowledgements}

We thank P.P.~Papadopoulos for enlightening discussions on the details
of the CR-regulated model for the IMF. FF and GDL acknowledge
financial support from the grants PRIN INAF 2014 ``Glittering
kaleidoscopes in the sky: the multifaceted nature and role of Galaxy
Clusters'' (1.05.01.94.02). GDL acknowledges financial support from
the MERAC foundation. MH acknowledges financial support from the
European Research Council via an Advanced Grant under grant agreement
no. 321323 (NEOGAL). GB acknowledges support for this work from the
National Autonomous University of M\'exico (UNAM), through grant
PAPIIT IG100115.

\bibliographystyle{mn2e}
\bibliography{fontanot}

\begin{thebibliography}{}

\bibitem[\protect\citeauthoryear{{Bruzual} \& {Charlot}}{{Bruzual} \&
  {Charlot}}{2003}]{Bruzual03}
{Bruzual} G.,  {Charlot} S.,  2003, \mnras, 344, 1000

\bibitem[\protect\citeauthoryear{{Cappellari}, {McDermid}, {Alatalo}, {Blitz},
  {Bois}, {Bournaud}, {Bureau}, {Crocker} \& et al.}{{Cappellari}
  et~al.}{2012}]{Cappellari12}
{Cappellari} M.,  {McDermid} R.~M.,  {Alatalo} K.,  {Blitz} L.,  {Bois} M.,
  {Bournaud} F.,  {Bureau} M.,  {Crocker} A.~F.,    et al. 2012, \nat, 484, 485

\bibitem[\protect\citeauthoryear{{Chabrier}}{{Chabrier}}{2003}]{Chabrier03}
{Chabrier} G.,  2003, \apjl, 586, L133

\bibitem[\protect\citeauthoryear{{Charlot} \& {Fall}}{{Charlot} \&
  {Fall}}{2000}]{CharlotFall00}
{Charlot} S.,  {Fall} S.~M.,  2000, \apj, 539, 718

\bibitem[\protect\citeauthoryear{{Conroy}, {Dutton}, {Graves}, {Mendel} \& {van
  Dokkum}}{{Conroy} et~al.}{2013}]{Conroy13}
{Conroy} C.,  {Dutton} A.~A.,  {Graves} G.~J.,  {Mendel} J.~T.,    {van Dokkum}
  P.~G.,  2013, \apjl, 776, L26

\bibitem[\protect\citeauthoryear{{Conroy} \& {van Dokkum}}{{Conroy} \& {van
  Dokkum}}{2012}]{ConroyvanDokkum12}
{Conroy} C.,  {van Dokkum} P.~G.,  2012, \apj, 760, 71

\bibitem[\protect\citeauthoryear{{De Lucia} \& {Blaizot}}{{De Lucia} \&
  {Blaizot}}{2007}]{DeLuciaBlaizot07}
{De Lucia} G.,  {Blaizot} J.,  2007, \mnras, 375, 2

\bibitem[\protect\citeauthoryear{{De Lucia}, {Fontanot} \& {Hirschmann}}{{De
  Lucia} et~al.}{2017}]{DeLucia17}
{De Lucia} G.,  {Fontanot} F.,    {Hirschmann} M.,  2017, \mnras, 466, L88

\bibitem[\protect\citeauthoryear{{De Lucia}, {Springel}, {White}, {Croton} \&
  {Kauffmann}}{{De Lucia} et~al.}{2006}]{DeLucia06}
{De Lucia} G.,  {Springel} V.,  {White} S.~D.~M.,  {Croton} D.,    {Kauffmann}
  G.,  2006, \mnras, 366, 499

\bibitem[\protect\citeauthoryear{{De Lucia}, {Tornatore}, {Frenk}, {Helmi},
  {Navarro} \& {White}}{{De Lucia} et~al.}{2014}]{DeLucia14}
{De Lucia} G.,  {Tornatore} L.,  {Frenk} C.~S.,  {Helmi} A.,  {Navarro} J.~F.,
    {White} S.~D.~M.,  2014, \mnras, 445, 970

\bibitem[\protect\citeauthoryear{{Elmegreen}, {Klessen} \&
  {Wilson}}{{Elmegreen} et~al.}{2008}]{Elmegreen08}
{Elmegreen} B.~G.,  {Klessen} R.~S.,    {Wilson} C.~D.,  2008, \apj, 681, 365

\bibitem[\protect\citeauthoryear{{Ferreras}, {La Barbera}, {de la Rosa},
  {Vazdekis}, {de Carvalho}, {Falc{\'o}n-Barroso} \& {Ricciardelli}}{{Ferreras}
  et~al.}{2013}]{Ferreras13}
{Ferreras} I.,  {La Barbera} F.,  {de la Rosa} I.~G.,  {Vazdekis} A.,  {de
  Carvalho} R.~R.,  {Falc{\'o}n-Barroso} J.,    {Ricciardelli} E.,  2013,
  \mnras, 429, L15

\bibitem[\protect\citeauthoryear{{Fontanot}}{{Fontanot}}{2014}]{Fontanot14}
{Fontanot} F.,  2014, \mnras, 442, 3138

\bibitem[\protect\citeauthoryear{{Fontanot}, {De Lucia}, {Benson}, {Monaco} \&
  {Boylan-Kolchin}}{{Fontanot} et~al.}{2013}]{Fontanot13}
{Fontanot} F.,  {De Lucia} G.,  {Benson} A.~J.,  {Monaco} P.,
  {Boylan-Kolchin} M.,  2013, ArXiv e-prints (arXiv:1301.4220)

\bibitem[\protect\citeauthoryear{{Fontanot}, {De Lucia}, {Hirschmann},
  {Bruzual}, {Charlot} \& {Zibetti}}{{Fontanot} et~al.}{2017}]{Fontanot17a}
{Fontanot} F.,  {De Lucia} G.,  {Hirschmann} M.,  {Bruzual} G.,  {Charlot} S.,
    {Zibetti} S.,  2017, \mnras, 464, 3812

\bibitem[\protect\citeauthoryear{{Fontanot}, {De Lucia}, {Monaco}, {Somerville}
  \& {Santini}}{{Fontanot} et~al.}{2009}]{Fontanot09b}
{Fontanot} F.,  {De Lucia} G.,  {Monaco} P.,  {Somerville} R.~S.,    {Santini}
  P.,  2009, \mnras, 397, 1776

\bibitem[\protect\citeauthoryear{{Fontanot}, {Hirschmann} \& {De
  Lucia}}{{Fontanot} et~al.}{2017}]{Fontanot17b}
{Fontanot} F.,  {Hirschmann} M.,    {De Lucia} G.,  2017, \apjl, 842, L14

\bibitem[\protect\citeauthoryear{{Goldsmith}}{{Goldsmith}}{2001}]{Goldsmith01}
{Goldsmith} P.~F.,  2001, \apj, 557, 736

\bibitem[\protect\citeauthoryear{{Gunawardhana}, {Hopkins}, {Sharp}, {Brough},
  {Taylor}, {Bland-Hawthorn}, {Maraston}, {Tuffs} \& et al.}{{Gunawardhana}
  et~al.}{2011}]{Gunawardhana11}
{Gunawardhana} M.~L.~P.,  {Hopkins} A.~M.,  {Sharp} R.~G.,  {Brough} S.,
  {Taylor} E.,  {Bland-Hawthorn} J.,  {Maraston} C.,  {Tuffs} R.~J.,    et al.
  2011, \mnras, 415, 1647

\bibitem[\protect\citeauthoryear{{Guo}, {White}, {Angulo}, {Henriques},
  {Lemson}, {Boylan-Kolchin}, {Thomas} \& {Short}}{{Guo} et~al.}{2013}]{Guo13}
{Guo} Q.,  {White} S.,  {Angulo} R.~E.,  {Henriques} B.,  {Lemson} G.,
  {Boylan-Kolchin} M.,  {Thomas} P.,    {Short} C.,  2013, \mnras, 428, 1351

\bibitem[\protect\citeauthoryear{{Guo}, {White}, {Boylan-Kolchin}, {De Lucia},
  {Kauffmann}, {Lemson}, {Li}, {Springel} \& {Weinmann}}{{Guo}
  et~al.}{2011}]{Guo11}
{Guo} Q.,  {White} S.,  {Boylan-Kolchin} M.,  {De Lucia} G.,  {Kauffmann} G.,
  {Lemson} G.,  {Li} C.,  {Springel} V.,    {Weinmann} S.,  2011, \mnras, 413,
  101

\bibitem[\protect\citeauthoryear{{Hennebelle} \& {Chabrier}}{{Hennebelle} \&
  {Chabrier}}{2008}]{HennebelleChabrier08}
{Hennebelle} P.,  {Chabrier} G.,  2008, \apj, 684, 395

\bibitem[\protect\citeauthoryear{{Henriques}, {White}, {Thomas}, {Angulo},
  {Guo}, {Lemson} \& {Springel}}{{Henriques} et~al.}{2013}]{Henriques13}
{Henriques} B.~M.~B.,  {White} S.~D.~M.,  {Thomas} P.~A.,  {Angulo} R.~E.,
  {Guo} Q.,  {Lemson} G.,    {Springel} V.,  2013, \mnras, 431, 3373

\bibitem[\protect\citeauthoryear{{Hirschmann}, {De Lucia} \&
  {Fontanot}}{{Hirschmann} et~al.}{2016}]{Hirschmann16}
{Hirschmann} M.,  {De Lucia} G.,    {Fontanot} F.,  2016, \mnras, 461, 1760

\bibitem[\protect\citeauthoryear{{Hopkins}}{{Hopkins}}{2012}]{Hopkins12}
{Hopkins} P.~F.,  2012, \mnras, 423, 2037

\bibitem[\protect\citeauthoryear{{Jasche}, {Ciardi} \& {En{\ss}lin}}{{Jasche}
  et~al.}{2007}]{Jasche07}
{Jasche} J.,  {Ciardi} B.,    {En{\ss}lin} T.~A.,  2007, \mnras, 380, 417

\bibitem[\protect\citeauthoryear{{Johansson}, {Thomas} \&
  {Maraston}}{{Johansson} et~al.}{2012}]{Johansson12}
{Johansson} J.,  {Thomas} D.,    {Maraston} C.,  2012, \mnras, 421, 1908

\bibitem[\protect\citeauthoryear{{Klessen}, {Ballesteros-Paredes},
  {V{\'a}zquez-Semadeni} \& {Dur{\'a}n-Rojas}}{{Klessen}
  et~al.}{2005}]{Klessen05}
{Klessen} R.~S.,  {Ballesteros-Paredes} J.,  {V{\'a}zquez-Semadeni} E.,
  {Dur{\'a}n-Rojas} C.,  2005, \apj, 620, 786

\bibitem[\protect\citeauthoryear{{Klessen}, {Spaans} \& {Jappsen}}{{Klessen}
  et~al.}{2007}]{Klessen07}
{Klessen} R.~S.,  {Spaans} M.,    {Jappsen} A.-K.,  2007, \mnras, 374, L29

\bibitem[\protect\citeauthoryear{{Kroupa}}{{Kroupa}}{2001}]{Kroupa01}
{Kroupa} P.,  2001, \mnras, 322, 231

\bibitem[\protect\citeauthoryear{{Kroupa} \& {Weidner}}{{Kroupa} \&
  {Weidner}}{2003}]{KroupaWeidner03}
{Kroupa} P.,  {Weidner} C.,  2003, \apj, 598, 1076

\bibitem[\protect\citeauthoryear{{Kroupa}, {Weidner}, {Pflamm-Altenburg},
  {Thies}, {Dabringhausen}, {Marks} \& {Maschberger}}{{Kroupa}
  et~al.}{2013}]{Kroupa13}
{Kroupa} P.,  {Weidner} C.,  {Pflamm-Altenburg} J.,  {Thies} I.,
  {Dabringhausen} J.,  {Marks} M.,    {Maschberger} T.,  2013, Planets, Stars
  and Stellar Systems.~Volume 5: Galactic Structure and Stellar Populations, 5,
  115

\bibitem[\protect\citeauthoryear{{Krumholz}}{{Krumholz}}{2014}]{Krumholz14}
{Krumholz} M.~R.,  2014, ArXiv e-prints (arXiv:1402.0867)

\bibitem[\protect\citeauthoryear{{La Barbera}, {Ferreras}, {Vazdekis}, {de la
  Rosa}, {de Carvalho}, {Trevisan}, {Falc{\'o}n-Barroso} \& {Ricciardelli}}{{La
  Barbera} et~al.}{2013}]{LaBarbera13}
{La Barbera} F.,  {Ferreras} I.,  {Vazdekis} A.,  {de la Rosa} I.~G.,  {de
  Carvalho} R.~R.,  {Trevisan} M.,  {Falc{\'o}n-Barroso} J.,    {Ricciardelli}
  E.,  2013, \mnras, 433, 3017

\bibitem[\protect\citeauthoryear{{La Barbera}, {Vazdekis}, {Ferreras},
  {Pasquali}, {Allende Prieto}, {R{\"o}ck}, {Aguado} \& {Peletier}}{{La
  Barbera} et~al.}{2017}]{LaBarbera17}
{La Barbera} F.,  {Vazdekis} A.,  {Ferreras} I.,  {Pasquali} A.,  {Allende
  Prieto} C.,  {R{\"o}ck} B.,  {Aguado} D.~S.,    {Peletier} R.~F.,  2017,
  \mnras, 464, 3597

\bibitem[\protect\citeauthoryear{{Leier}, {Ferreras}, {Saha}, {Charlot},
  {Bruzual} \& {La Barbera}}{{Leier} et~al.}{2016}]{Leier16}
{Leier} D.,  {Ferreras} I.,  {Saha} P.,  {Charlot} S.,  {Bruzual} G.,    {La
  Barbera} F.,  2016, \mnras, 459, 3677

\bibitem[\protect\citeauthoryear{{Marigo}, {Girardi}, {Bressan}, {Groenewegen},
  {Silva} \& {Granato}}{{Marigo} et~al.}{2008}]{Marigo08}
{Marigo} P.,  {Girardi} L.,  {Bressan} A.,  {Groenewegen} M.~A.~T.,  {Silva}
  L.,    {Granato} G.~L.,  2008, \aap, 482, 883

\bibitem[\protect\citeauthoryear{{Marks}, {Kroupa}, {Dabringhausen} \&
  {Pawlowski}}{{Marks} et~al.}{2012}]{Marks12}
{Marks} M.,  {Kroupa} P.,  {Dabringhausen} J.,    {Pawlowski} M.~S.,  2012,
  \mnras, 422, 2246

\bibitem[\protect\citeauthoryear{{McWilliam}, {Wallerstein} \&
  {Mottini}}{{McWilliam} et~al.}{2013}]{McWilliam13}
{McWilliam} A.,  {Wallerstein} G.,    {Mottini} M.,  2013, \apj, 778, 149

\bibitem[\protect\citeauthoryear{{Muratov}, {Kere{\v s}},
  {Faucher-Gigu{\`e}re}, {Hopkins}, {Quataert} \& {Murray}}{{Muratov}
  et~al.}{2015}]{Muratov15}
{Muratov} A.~L.,  {Kere{\v s}} D.,  {Faucher-Gigu{\`e}re} C.-A.,  {Hopkins}
  P.~F.,  {Quataert} E.,    {Murray} N.,  2015, \mnras, 454, 2691

\bibitem[\protect\citeauthoryear{{Papadopoulos}}{{Papadopoulos}}{2010}]{Papadopoulos10}
{Papadopoulos} P.~P.,  2010, \apj, 720, 226

\bibitem[\protect\citeauthoryear{{Papadopoulos} \& {Thi}}{{Papadopoulos} \&
  {Thi}}{2013}]{Papadopoulos13}
{Papadopoulos} P.~P.,  {Thi} W.-F.,  2013, in {Torres} D.~F.,  {Reimer} O.,
  eds, Cosmic Rays in Star-Forming Environments Vol.~34 of Astrophysics and
  Space Science Proceedings, {The Initial Conditions of Star Formation: Cosmic
  Rays as the Fundamental Regulators}.
p.~41

\bibitem[\protect\citeauthoryear{{Papadopoulos}, {Thi}, {Miniati} \&
  {Viti}}{{Papadopoulos} et~al.}{2011}]{Papadopoulos11}
{Papadopoulos} P.~P.,  {Thi} W.-F.,  {Miniati} F.,    {Viti} S.,  2011, \mnras,
  414, 1705

\bibitem[\protect\citeauthoryear{{Planck Collaboration XVI}}{{Planck
  Collaboration XVI}}{2014}]{Planck_cosmpar}
{Planck Collaboration XVI} 2014, \aap, 571, A16

\bibitem[\protect\citeauthoryear{{Romano}, {Matteucci}, {Zhang}, {Papadopoulos}
  \& {Ivison}}{{Romano} et~al.}{2017}]{Romano17}
{Romano} D.,  {Matteucci} F.,  {Zhang} Z.-Y.,  {Papadopoulos} P.~P.,
  {Ivison} R.~J.,  2017, \mnras, 470, 401

\bibitem[\protect\citeauthoryear{{Salpeter}}{{Salpeter}}{1955}]{Salpeter55}
{Salpeter} E.~E.,  1955, \apj, 121, 161

\bibitem[\protect\citeauthoryear{{Smith}}{{Smith}}{2014}]{Smith14}
{Smith} R.~J.,  2014, \mnras, 443, L69

\bibitem[\protect\citeauthoryear{{Smith}, {Lucey} \& {Conroy}}{{Smith}
  et~al.}{2015}]{Smith15}
{Smith} R.~J.,  {Lucey} J.~R.,    {Conroy} C.,  2015, \mnras, 449, 3441

\bibitem[\protect\citeauthoryear{{Spiniello}, {Trager}, {Koopmans} \&
  {Chen}}{{Spiniello} et~al.}{2012}]{Spiniello12}
{Spiniello} C.,  {Trager} S.~C.,  {Koopmans} L.~V.~E.,    {Chen} Y.~P.,  2012,
  \apjl, 753, L32

\bibitem[\protect\citeauthoryear{{Springel}, {White}, {Jenkins}, {Frenk},
  {Yoshida}, {Gao}, {Navarro}, {Thacker}, {Croton}, {Helly}, {Peacock}, {Cole},
  {Thomas}, {Couchman}, {Evrard}, {Colberg} \& {Pearce}}{{Springel}
  et~al.}{2005}]{Springel05}
{Springel} V.,  {White} S.~D.~M.,  {Jenkins} A.,  {Frenk} C.~S.,  {Yoshida} N.,
   {Gao} L.,  {Navarro} J.,  {Thacker} R.,  {Croton} D.,  {Helly} J.,
  {Peacock} J.~A.,  {Cole} S.,  {Thomas} P.,  {Couchman} H.,  {Evrard} A.,
  {Colberg} J.,    {Pearce} F.,  2005, \nat, 435, 629

\bibitem[\protect\citeauthoryear{{Thi}, {van Dishoeck}, {Bell}, {Viti} \&
  {Black}}{{Thi} et~al.}{2009}]{Thi09}
{Thi} W.-F.,  {van Dishoeck} E.~F.,  {Bell} T.,  {Viti} S.,    {Black} J.,
  2009, \mnras, 400, 622

\bibitem[\protect\citeauthoryear{{Thomas}, {Maraston}, {Schawinski}, {Sarzi} \&
  {Silk}}{{Thomas} et~al.}{2010}]{Thomas10}
{Thomas} D.,  {Maraston} C.,  {Schawinski} K.,  {Sarzi} M.,    {Silk} J.,
  2010, \mnras, 404, 1775

\bibitem[\protect\citeauthoryear{{Tortora}, {La Barbera} \&
  {Napolitano}}{{Tortora} et~al.}{2016}]{Tortora16}
{Tortora} C.,  {La Barbera} F.,    {Napolitano} N.~R.,  2016, \mnras, 455, 308

\bibitem[\protect\citeauthoryear{{Wang}, {De Lucia}, {Kitzbichler} \&
  {White}}{{Wang} et~al.}{2008}]{Wang08}
{Wang} J.,  {De Lucia} G.,  {Kitzbichler} M.~G.,    {White} S.~D.~M.,  2008,
  \mnras, 384, 1301

\bibitem[\protect\citeauthoryear{{Weidner} \& {Kroupa}}{{Weidner} \&
  {Kroupa}}{2005}]{WeidnerKroupa05}
{Weidner} C.,  {Kroupa} P.,  2005, \apj, 625, 754

\bibitem[\protect\citeauthoryear{{Xie}, {De Lucia}, {Hirschmann}, {Fontanot} \&
  {Zoldan}}{{Xie} et~al.}{2017}]{Xie17}
{Xie} L.,  {De Lucia} G.,  {Hirschmann} M.,  {Fontanot} F.,    {Zoldan} A.,
  2017, \mnras, 469, 968

\bibitem[\protect\citeauthoryear{{Yan}, {Jerabkova} \& {Kroupa}}{{Yan}
  et~al.}{2017}]{Yan17}
{Yan} Z.,  {Jerabkova} T.,    {Kroupa} P.,  2017, \aap, 607, A126

\bibitem[\protect\citeauthoryear{{Zibetti}, {Charlot} \& {Rix}}{{Zibetti}
  et~al.}{2009}]{Zibetti09}
{Zibetti} S.,  {Charlot} S.,    {Rix} H.,  2009, \mnras, 400, 1181

\end{thebibliography}

\appendix

\section{Model Calibration}\label{app:recal}
\begin{figure*}
  \centerline{ \includegraphics[width=18cm]{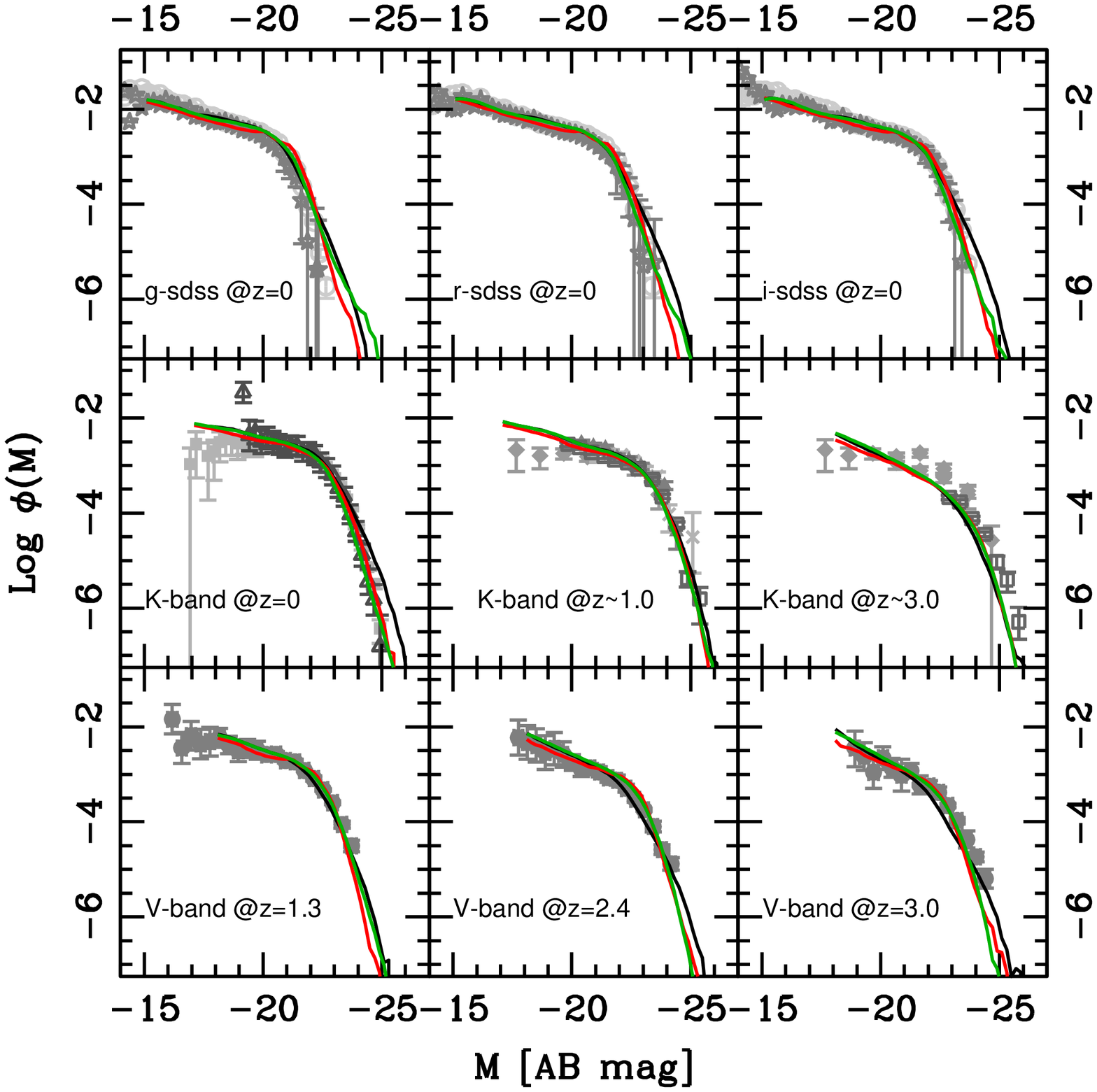} }
  \caption{Model calibration set. Predicted luminosity functions in
    different wavebands and at different redshifts. Solid Black, red
    and green lines refer to the H16F, IGIMF and PP11 runs
    respectively. Grey points show the compilation of observational
    estimates in the SDSS g, r and i-band, K- and V-band from F17a
    (see detailed references herein).}\label{fig:calibration}
\end{figure*}
The original H16F model is tuned against the reconstructed evolution
of physical quantities such as $M_\star$ and $SFR$. However, such
constraints are usually derived under the hypothesis of a universal
IMF and therefore cannot be used in a variable IMF scenario. As
already discussed in F17a, our model, based on the PP11 approach,
requires a recalibration of key model parameters, given the different
impact on galaxy evolution of the assumed IMFs (in terms of baryons
locked in long-living stars, fraction of SNe and ratio between SNIa
and SNII). In detail, the relevant parameters are the SFR efficiency
($\alpha_{\rm SF}$), AGN feedback ($\kappa_{\rm radio}$), stellar
feedback reheating ($\epsilon_{\rm reheat}$) and ejection rate
($\epsilon_{\rm eject}$), and the reincorporation rate ($\gamma_{\rm
  reinc}$). We refer the interested reader to \citet{Hirschmann16} for
a more detailed discussion of the role these parameters play in the
{\gaea} context. The assumed calibration set is the same defined in
F17a (see there for a full reference list), i.e. the evolution of the
$K$- and $V$-band LF at $z \lesssim 3$, and the $z=0$ LFs in the Sloan
Digital Sky Survey (SDSS) $g$, $r$ and
$i$-bands. Fig.~\ref{fig:calibration} compares the resulting
luminosity functions (green lines) with those obtained in the context
of the IGIMF theory (red lines - High-$\alpha_{\rm SF}$ model in F17a)
and using the reference H16F model (black lines).

\section{Physical properties of model galaxies}\label{app:phys}
\begin{figure}
  \centerline{ \includegraphics[width=9cm]{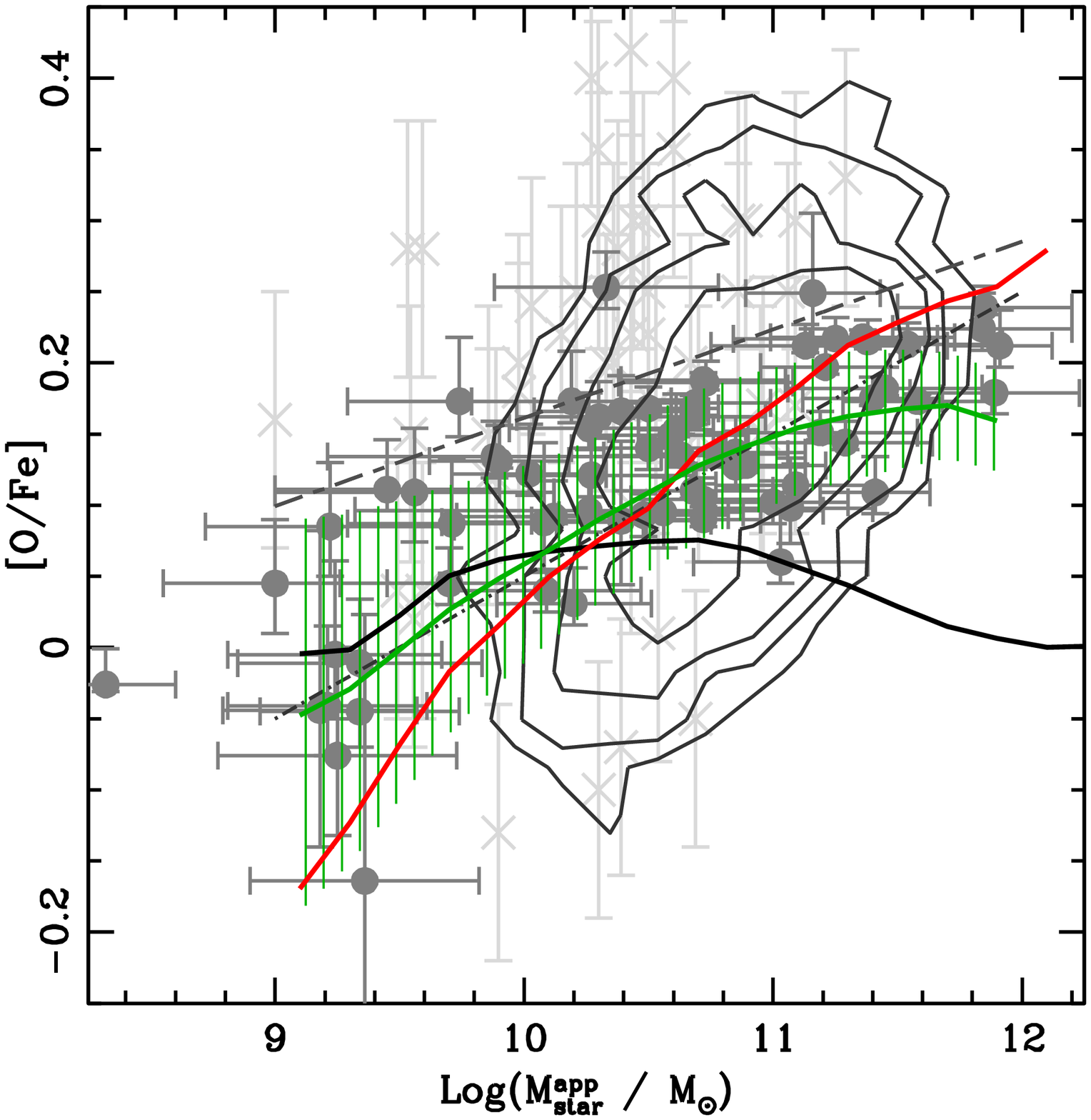} }
  \caption{Predicted [O/Fe] ratios in $B/T>0.7$ model galaxies
    compared to the observed [$\alpha$/Fe] ratios in local elliptical
    galaxies. Data collection (grey points, contours, dot-dashed and
    long-short dashed lines) as in F17a (see detailed references
    herein). }\label{fig:enhanced}
\end{figure}
\begin{figure}
  \centerline{ \includegraphics[width=9cm]{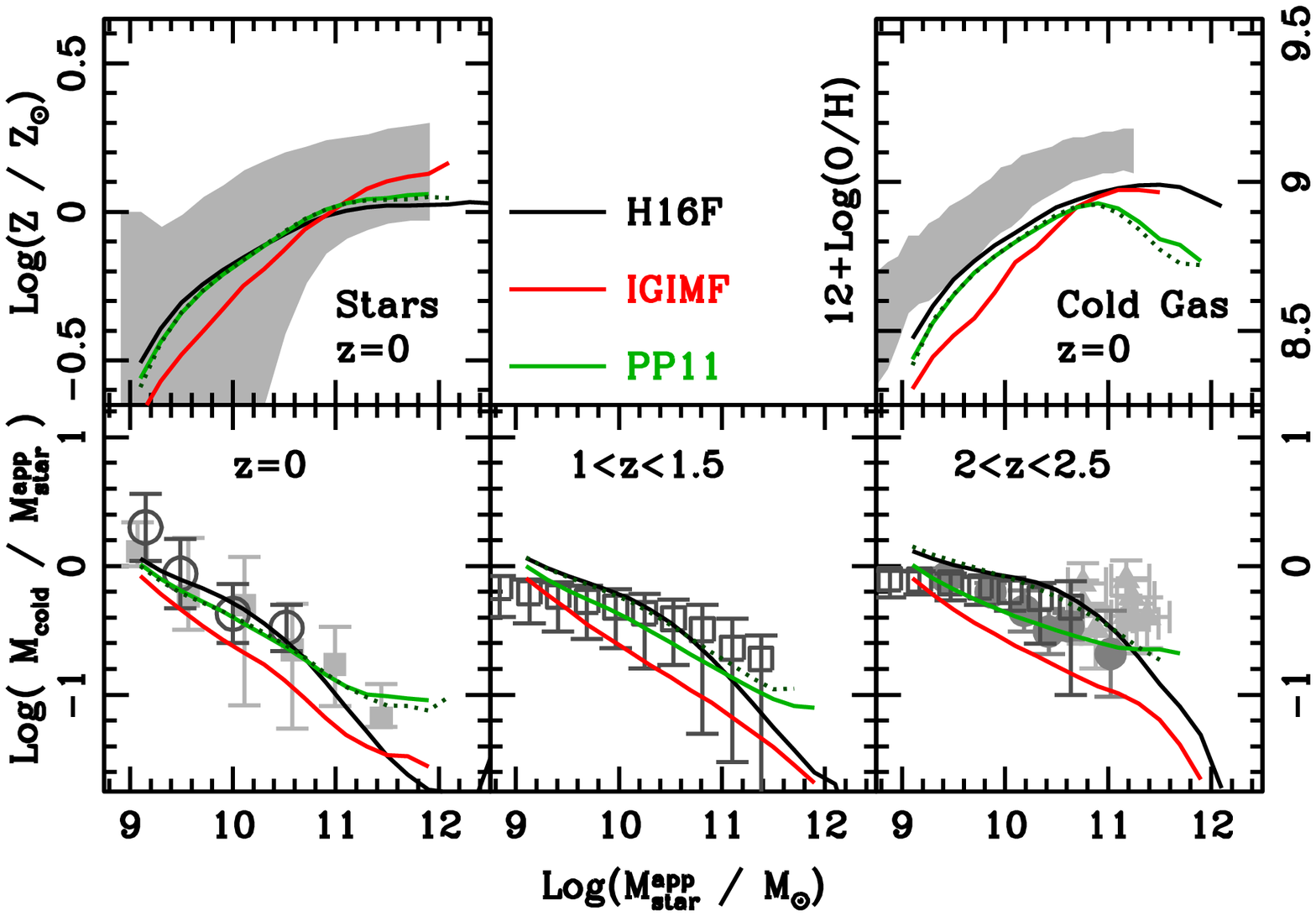} }
  \caption{Evolution of key galaxy properties as a function of
    photometrically estimated $M^{\rm app}_{\star}$. {\it Bottom
      panel:} evolution of the cold gas fraction in star-forming
    galaxies (i.e. SFR/$M^{\rm app}_{\star} > 10^{-2}$ Gyrs$^{-1}$);
    {\it left upper panel:} $z=0$ total stellar metallicity; {\it
      right upper panel:} $z=0$ cold gas metallicity
    metallicity. Lines and colours as in
    Fig.~\ref{fig:calibration}. In all panels, grey points and areas
    represents observational constraints as in F17a (see references
    herein).}\label{fig:phyprop}
\end{figure}
\begin{figure*}
  \centerline{ \includegraphics[width=9cm]{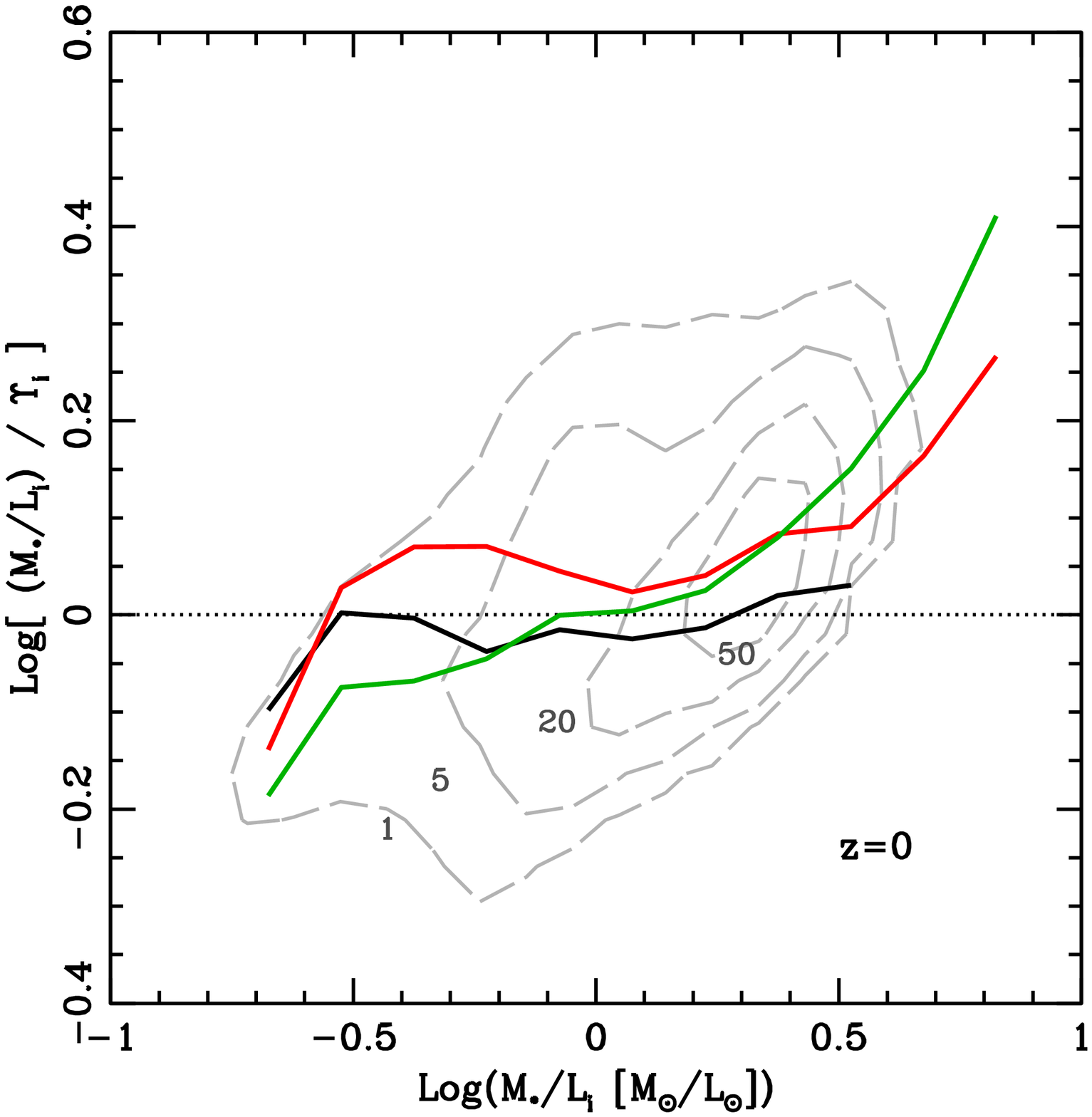}
    \includegraphics[width=9cm]{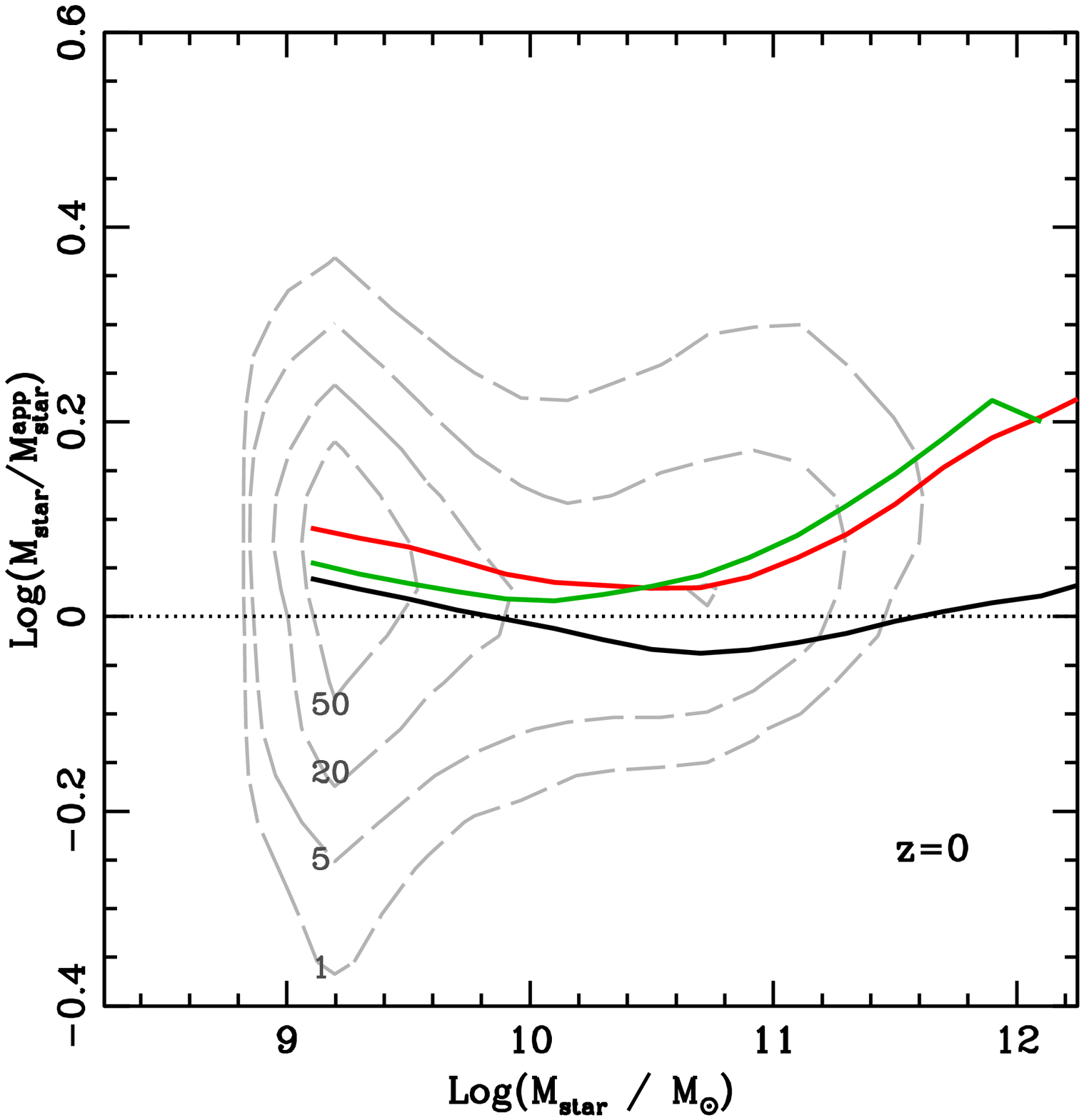} }
  \caption{Left panel: ratio of proper $M_\star/L_i$ and the
    photometric equivalent $\Upsilon_i$, as a function of
    $M_\star/L_i$. Right panel: $M_\star/M_\star^{\rm app}$ versus
    $M_\star$. In each panel, only $B/T>0.7$ galaxies have been
    considered and colours and linetypes corresponds as in
    Fig.~\ref{fig:calibration}. Grey dashed contours mark galaxy
    number densities (in percentage of the maximum density) in the
    PP11 run.}\label{fig:demas}
\end{figure*}
In this Appendix we present the physical properties of model in the
PP11 run, and contrast them against the H16F and IGIMF results. In
all plots the green, red and black solid lines refer to the the PP11,
IGIMF and H16F runs respectively. Unless differently stated, we always
show predictions as a function of the apparent stellar mass ($M^{\rm
  app}_{\star}$), computed from synthetic magnitudes using a
mass-to-light vs colour relation as routinely done in observational
samples \citep[see e.g.][]{Zibetti09}. In detail, we adopt a relation
proposed by Zibetti et al. (in prep.):

\begin{equation}\label{eq:zibetti}
\log \Upsilon_i = \upsilon (g-i) +\delta +\epsilon
\end{equation} 

\noindent
where $\Upsilon_i$ represents the stellar mass-to-light ratio in the
$i$-band. The best-fit coefficients $\upsilon=0.9$ and $\delta=0.7$
are derived using a Monte Carlo library of 500,000 synthetic stellar
population spectra, based on the revised version of the
\citet{Bruzual03} simple stellar population tracks, and using the
age-dependent dust attenuation prescription from
\citet{CharlotFall00}. Based on F17a results, we also include the
additional factor $\epsilon=0.13$ in order to account for a shift in
$M_\star^{\rm app}$ with respect to $M_\star$, due to spatial
resolution effects (see F17a for a complete discussion on the origin
of this shift). This choice ensure that $M_\star^{\rm app}$ and
$M_\star$ are statistically equivalent in the H16F run (modulo some
scatter). Eq.~\ref{eq:zibetti} provides a Chabrier (2003) IMF
equivalent stellar mass; in the following, we neglect the small
difference in normalization between the \citet[which better describes
  the general form of the PP11 IMFs]{Kroupa01} and the
\citet{Chabrier03} IMF.

Overall, we find similar trends in the physical properties of model
galaxies predicted in the IGIMF and PP11 runs. The consistencies
between the two approaches are particularly interesting, since they
predict similar behaviour of the IMF in strongly star forming sources
(either via high-SFR or high-$\Sigma_{\rm SFR}$), i.e. a top-heavier IMF
with respect to the local neighbourhood, but with a different shape
evolution. In the IGIMF theory, the main evolution of the IMF is
predicted at its high-mass end, while the low-mass end and the
low-mass characteristic mass are unaffected. In PP11, both slopes are
fixed, and the effect of a different $U_{\rm CR}$ is mainly seen as a
change in the characteristic mass.

However, some interesting differences are seen, which could be in
principle used to discriminate between the two approaches. In
Fig.~\ref{fig:enhanced}, we consider the relation between
[$\alpha$/Fe] enrichment and stellar mass for $B/T>0.7$ galaxies
(which are considered a good proxy for elliptical galaxies). As in
F17a, we will refer to [$\alpha$/Fe] ratios for the observational data
and to the [O/Fe] for theoretical predictions. This choice is related
to the fact that oxygen represents the most abundant (so a good
tracer) among $\alpha$-elements, even if most observational estimate
are actually derived from Magnesium lines. The predicted slope (green
solid line - the hatched area represents the 1-$\sigma$ scatter) of
the [$\alpha$/Fe]-stellar mass relation in PP11 is closer to the fits
to the data by \citet{Thomas10} and \citet{Johansson12}, while IGIMF
theory predicts a somehow steeper relation, but still compatible with
the data.

In Fig.\ref{fig:phyprop}, we present a collection of the additional
relevant physical properties in galaxy evolution studies, namely the
evolution of the cold gas fractions of star-forming galaxies
(SFR/$M^{\rm app}_{\star} > 10^{-2}$ Gyr$^{-1}$ - bottom panels) and
$z=0$ mass-metallicity relations (total metallicity - upper left panel
- and cold gas metallicity - upper right panel). Cold gas fractions
(and their redshift evolution) appear to be the direct observable most
sensitive to IMF variation. Nonetheless, the present uncertainties on
observed samples prevent a clear separation between models: PP11
better reproduce the observational data, but the IGIMF run is still
within the 1-$\sigma$ scatter. It is also worth pointing out the
peculiar behaviour of PP11 predictions on the cold gas
mass-metallicity relation (Fig.~\ref{fig:phyprop} - upper right
panel). The model predict a clear turnover at $\sim 10^{11} \msun$,
which is slightly outside the range covered by the data. It is however
unclear if this feature could be used as a test for a variable IMF
model, since we test that this is particularly sensible to the
strength of AGN feedback and its interplay with stellar feedback
schemes.

In Fig.~\ref{fig:demas} we show the ratio between photometrically
derived and intrinsic quantities as a function of intrinsic quantities
for $B/T>0.7$. Those plots are meant to be compared with the
corresponding observational data suggesting deviations from the
assumption of a universal (Chabrier-like) IMF. The left panel
considers the ratio of proper stellar mass-to-light ratios in the
$i$-band ($M_\star/L_i$) and the photometric equivalent $\Upsilon_i$
from Eq.~\ref{eq:zibetti}: this plot has to be compared to the
dynamical analysis from \citealt{Cappellari12} in the ATLAS$^{3D}$
sample of early-type galaxies. The right panel shows the evolution of
the $M_\star/M_\star^{\rm app}$ ratio as a function of the intrinsic
stellar mass and roughly corresponds to the spectroscopic analysis as
in \citet{Conroy13}. Interesting differences between the IGIMF and
PP11 runs are also evident in these plots and are particularly
relevant for the dynamical estimate (Fig.~\ref{fig:demas}, right
panel), with the PP11 runs showing a clear steeping of the relation
with respect to the IGIMF run. The steeping is particularly relevant
at low-mass-to-light ratios, where the PP11 model predicts a {\it
  deficit} in the photometric with respect to the dynamical estimate.

\end{document}